\documentclass[draftcls, 12pt, onecolumn]{IEEEtran}
\usepackage{bbm}
\usepackage{mathrsfs}
\usepackage{amssymb}
\usepackage{bbding}
\usepackage{amsfonts}
\ifCLASSINFOpdf

\else

\fi

\usepackage{graphicx,cite,epsfig,amssymb,amsmath,multirow,extarrows}
\usepackage{clrscode}
\usepackage{caption}
\usepackage{algorithm}
\usepackage{algpseudocode}
\usepackage{algpascal}
\usepackage{multirow}
\usepackage{amsmath}
\usepackage{xcolor}
\usepackage{bm}
\usepackage{hyperref}
\usepackage{amsfonts}
\usepackage{graphicx}
\usepackage{cite}
\usepackage{subfigure}
\usepackage{mathtools}
\usepackage{diagbox}
\usepackage[square, comma, sort&compress, numbers]{natbib}
\begin{document}

\title{Pilot Power Allocation Through User Grouping \\in Multi-Cell Massive MIMO Systems}

\author{Pei Liu, Shi Jin, \IEEEmembership{Member,~IEEE,} Tao Jiang, \IEEEmembership{Senior Member,~IEEE,}\\  Qi Zhang, and Michail Matthaiou, \IEEEmembership{Senior Member,~IEEE}
\thanks{P. Liu and T. Jiang are with Wuhan National Laboratory for Optoelectronics, School
of Electrical Information and Communications, Huazhong University of Science and Technology, Wuhan 430074,
China  (e-mail: peil@hust.edu.cn; Tao.Jiang@ieee.org).}
\thanks{S. Jin is with the National Mobile Communications Research Laboratory,
Southeast University, Nanjing 210096, China (e-mail: jinshi@seu.edu.cn).}
\thanks{Q. Zhang is with Jiangsu Key Laboratory of Wireless Communications, Nanjing University of Posts and Telecommunications, Nanjing 210003, China (e-mail: zhangqiqi\_1212@126.com).}
\thanks{M. Matthaiou is with the School of Electronics, Electrical Engineering and
Computer Science, Queen's University Belfast, Belfast, BT3 9DT, U.K. (e-mail: m.matthaiou@qub.ac.uk).}
}
\markboth{}
{}

\maketitle

\begin{abstract}
In this paper, we propose a relative channel estimation error (RCEE) metric, and derive closed-form expressions for its expectation $\rm {Exp}_{rcee}$ and the achievable uplink rate holding for any number of base station antennas $M$, with the least squares (LS) and minimum mean squared error (MMSE) estimation methods. It is found that  RCEE and $\rm {Exp}_{rcee}$ converge to the same constant value when $M\rightarrow\infty$, resulting in the pilot power allocation (PPA) is substantially simplified and a PPA algorithm is proposed to minimize the average $\rm {Exp}_{rcee}$ per user with a total pilot power budget $P$ in multi-cell massive multiple-input multiple-output systems. Numerical results show that the PPA algorithm brings considerable gains for the LS estimation compared with equal PPA (EPPA), while the gains are only significant with large frequency reuse factor (FRF) for the MMSE estimation. Moreover, for large FRF and large $P$, the performance of the LS approaches to the performance of the MMSE, which means that simple LS estimation method is a very viable when co-channel interference is small. For the achievable uplink rate, the PPA scheme delivers almost the same average achievable uplink rate and improves the  minimum achievable uplink rate compared with the EPPA scheme.
\end{abstract}


\IEEEpeerreviewmaketitle

\section{Introduction}

\IEEEPARstart{M}{ultiple-input} multiple-output (MIMO) systems have been well integrated in the fourth generation mobile communication technology since they can
improve high data rates and suppress channel fading effects. Unlike traditional MIMO, massive MIMO \cite{Marzetta10,Rusek13,QiZhang14} technology deploys  hundreds of antennas to serve tens of users who share the same time-frequency resources, and has attracted wide attention from academia and industry in recent years. Massive MIMO can achieve considerable spatial multiplexing gains and improve energy efficiency more effectively. Hence, it is considered as one of the key technologies for the fifth generation  mobile communication networks \cite{Boccardi14}.

A great amount of research results have been reported on massive MIMO, in the context of
linear precoders and detectors \cite{Ngo13,Jose11}, achievable sum-rate analysis \cite{QiZhang14, Hoydis13},
hardware impairments \cite{Bjornson14}, channel estimation \cite{Yin13,JiankangZhang14}, and so on. Among these important topics, based on the seminal work \cite{Marzetta10}, the channel estimation is one of the biggest challenges pertaining to massive MIMO. In this context, the time-division duplex (TDD) mode dominates the massive MIMO literature, since the pilot sequence length is analogous to the number of users \cite{Marzetta15}.
Interestingly, channel estimation performance can be quantified via different evaluation metrics. For example, the symbol error probability \cite{Ngo12} and the bit error rate  \cite{Muller14,Ma14} all reveal the channel estimation performance via the number of errors in the data transmission. The mean squared error (MSE) \cite{Wen15,Shariati14} discloses the quality of the estimated channel based on the absolute channel estimation error. In \cite{Shariati14}, a general MSE expression was given based on the  correlation matrix of the user channel. Considering the normalized relative error of the estimated channel, \cite{JiankangZhang14,Shariati14} utilized the normalized MSE (NMSE) metric, whilst \cite{Yin13} utilized the normalized channel estimation error metric. However, closed-form expressions for these normalized relative error metrics are barely available in the literature. Very recently, \cite{Hu15} provided an analytical expression for the NMSE, while mainly paying attention on how to extract the desired source data from the received signal; it also proposed a semi-blind channel estimation method. Finally, \cite{Zhu15} studied a channel estimation MSE metric, which involves the channel estimation error and the true channel, and compared the proposed soft pilot reuse scheme and the all pilot reuse scheme through simulations. From the above discussion, it becomes apparent that a detailed theoretical analysis of such normalized channel estimation error metrics for massive MIMO is missing from the open literature.

On a parallel avenue, pilot power allocation (PPA) represents another formidable opportunity in massive MIMO systems to improve the system performance. In \cite{Ngo14}, a resource allocation scheme was proposed to maximize the sum spectral efficiency (SE), which determines the optimal values of the pilot sequence length, the pilot signal power, and the data signal power. A joint pilot and data transmit power control was provided in \cite{Guo14} to minimize the total power consumption of all users with the constraints of per user signal to interference plus noise ratio (SINR) and per user power. In \cite{Xiang14}, a PPA policy was developed to maximize the minimum asymptotic SINR in each cell, by adopting a pilot scheme where all users in each cell share the same pilot sequence and keep the pilot sequences orthogonal for different cells. In \cite{Zheng14}, a binary search PPA algorithm was provided to  maximize the achievable downlink sum rate with matched filter based precoding in single cell massive MIMO. A pilot power control scheme was provided in \cite{Saxena15} to mitigate the pilot contamination (PC) effect \cite{Marzetta10,Jose11} by reducing the transmit power of users that are close to their target BSs. Recently, in \cite{Cheng15}, a joint optimal pilot and data power allocation was proposed in single cell uplink massive MIMO systems to maximize the minimum SE and maximize the sum SE through geometric programming and an algorithm based on Karush-Kuhn-Tucker  points, respectively.

In our work, we take a substantially different approach to the problem of channel estimation. In particular, we strive to improve channel estimation performance by designing a PPA scheme in a heuristic manner.  In this paper, we study the  performance of channel estimation and achievable uplink rate in  multi-cell massive MIMO systems based on the least squares (LS) and minimum mean squared error (MMSE) estimation methods, respectively, by considering a modified NMSE metric, called the relative channel estimation error (RCEE). Closed-form expressions for the expectation of RCEE ($\rm {Exp}_{rcee}$) and  achievable uplink rate are obtained, which depend only on the large-scale fading coefficients. We conclude that when the number of BS antennas ($M$) grows to infinity, the RCEE and the $\rm {Exp}_{rcee}$ approach to the some constant value. This important observation, which is due to the channel hardening effect, enables us to design a simple PPA algorithm, which aims at minimizing the average  $\rm {Exp}_{rcee}$ per user in the target cell. Numerical simulations justify the accuracy of our analytical results, and it is found that the proposed PPA algorithm can approach the solution of the general constrained optimization problem very effectively. For the channel estimation performance, compared with the equal PPA (EPPA) scheme, the MMSE PPA scheme obtains an increasing gain as the frequency reuse factor grows. However, the gain of the LS PPA scheme remains fixed. Most importantly, the simple LS estimation method can achieve almost the same performance with the MMSE estimation method, with both higher frequency reuse factors and larger total pilot power. For the achievable uplink rate performance, compared with the EPPA scheme, the PPA scheme maintains almost the same average achievable uplink rate and improves considerably the minimum achievable uplink rate.

The rest of the paper is organized as follows. The channel model and the procedure of the channel estimation and uplink data transmission are described in Section II. The optimization problem, the analysis of the RCEE and the $\rm { {Exp}}_{ {rcee}}$, and the corresponding PPA algorithm are presented in Section III. Section IV presents the simulation results to check the effectiveness of the PPA algorithm and compare the PPA and EPPA schemes on the basis of channel estimation and achievable uplink rate performance. Conclusions are presented in Section V.

\textit {Notation:} Lower-case (underlined lower-case) and upper-case (underlined upper-case) boldface letters denote vectors and matrices, respectively. The ${\mathbb{C}}^{M\times N}$ denotes the $M\times N$ complex space. The notations  ${\bf A}^{\text H}$, ${\bf A}^{-1}$ and ${\rm {tr}}({\bf A})$ indicate the Hermitian transpose, the inverse and the trace of the matrix ${\bf A}$, respectively. The $M\times M$ identity matrix is ${\bf I}_{M}$. The $M\times N$ zero matrix is ${\bf 0}_{M\times N}$. The expectation operation is $\mathbb{E}\{\cdot\}$. A complex Gaussian random vector ${\bf x}$ is denoted as ${\bf x}\sim\mathcal {C}\mathcal {N}(\bar{{\bf x}},{\bf {\Sigma}})$, where the mean vector is $\bar{{\bf x}}$ and the covariance matrix is ${\bf {\Sigma}}$. Finally, ${\rm {diag}}(\bf a)$ denotes a diagonal matrix where the main diagonal entries are the elements of $\bf a$, and $\|\cdot\|_{2}$ denotes the 2-norm of vector.

\section{System Model}
\subsection{Signal and Channel Model}

The system architecture is a typical cellular communication system with $L$ hexagonal cells. Each cell contains $K$ single-antenna users and one $M$-antenna BS. We assume that the BSs and users in the whole system are strictly synchronized. The whole system operates under a TDD protocol and adopts the orthogonal frequency division multiplexing (OFDM) technique. Moreover, for convenience, the dependency of the user's channel on the sub-carrier index is suppressed. Hence, for each channel use, the received signal vector ${\bf y}_{j}\in \mathbb{C}^{M\times1}$ at the BS in  cell $j$ is given by
\begin{align}
   {\bf y}_{j}=\sqrt{p_{u}}\sum_{l=1}^{L}{\bf H}_{jl}{\bf x}_{l}+{\bf n}_{j},\tag{1}
\end{align}
where ${\bf x}_{l}\in\mathbb{C}^{K\times 1}$ is the transmit signal vector of the users in  cell $l$, $p_{u}$ denotes the average normalized transmitted power of all users in all cells, and ${\bf n}_{j}\in \mathbb{C}^{M\times1}$ is the noise vector containing independent elements $\sim\mathcal {C}\mathcal {N}(0,1)$ in cell $j$. Also, ${\bf H}_{jl}\in\mathbb{C}^{M \times K}$ denotes the channel between all  users in cell $l$ and the BS in cell $j$, which is given by
\begin{align}
{\bf H}_{jl}\triangleq[{\bf h}_{jl1},\ldots,{\bf h}_{jlk} ,\ldots,{\bf h}_{jlK}],\tag{2}
\end{align}
where ${\bf h}_{jlk}$ is the uplink channel transmission vector between the user $k$ in cell $l$ and the BS in cell $j$. The channel ${\bf h}_{jlk}$ is modeled as a combination of small-scale fading and large-scale fading and is written as \cite{Marzetta10,Ngo13}
\begin{align}
   {\bf h}_{jlk}=\underline{\bf h}_{jlk}\beta_{jlk}^{\frac{1}{2}},\tag{3}
\end{align}
where $\underline{\bf h}_{jlk}\sim\mathcal {C}\mathcal {N}({\bf 0}_{M\times 1},{\bf I}_{M})$ is the small-scale fading and $\beta_{jlk}$ is the large-scale fading coefficient.
Here, we make the block fading assumption that the large-scale fading coefficients are kept fixed over lots of coherence time intervals and also assume that large-scale fading coefficients are known at the BS \cite{Ngo13}, while small-scale fading fading coefficients remain fixed within a coherence time interval. At the same time, each user's channel is considered to be independent from other users' channels.

\subsection{Channel Estimation}
At the start of the coherence interval, before the user sends data to the BS, the BS needs to acquire channel state information (CSI) by estimating the channel between the user and itself. We use uplink pilot sequences to perform channel estimation. The pilot sequence sent by the user $k$ in cell $l$ is
\begin{align}
   {\bf s}_{lk}&=\sqrt{\rho_{lk}}\underline{{\bf s}}_{lk}\in \mathbb{C}^{\tau\times 1},\tag{4}
\end{align}
where ${\underline{\bf s}}_{lk}\in \mathbb{C}^{\tau\times 1}$ is the  pilot sequence with length $\tau$ sent by the user $k$ in cell $l$, and ${\rho}_{lk}$ is the pilot power of the user $k$ in cell $l$.\footnote{Since we elaborate on the channel estimation performance, we allow the pilot power in (4) to be variable for different users in each cell, while the transmit data power in (1) is kept the same for all users in all cells. This assumption also offers analytical tractability.} Moreover,
to ensure the orthogonality of users' pilot sequences within one cell, we set the pilot sequence length to be larger than the number of users, i.e., $\tau\ge K$, and
\begin{align}
{\underline{\bf s}}_{lk}^{\text H}{\underline{\bf s}}_{lk}=1 \ \text {and}\  {\underline{\bf s}}_{lk_1}^{\text H}{\underline{\bf s}}_{lk_2}=0,\ \forall k_1\neq k_2.\tag{5}
\end{align}

From the perspective of PC, the worst case choice is to  reuse ${\underline{\bf s}}_{lk}$  in all $L$ cells \cite{Marzetta10,Jose11,Yin13} for user $k$. That is, $\forall i,j\in \{1, \ldots , L\}$,
\begin{align}
{\underline{\bf s}}_{lk}={\underline{\bf s}}_{jk}.\tag{6}
\end{align}
Hence, in the phase of channel estimation, the BS in cell $j$ receives the signal matrix
\begin{align}
   {\bf Y}_{j}&=\sum_{l=1}^{L}\sum_{k=1}^{K}\sqrt{\rho_{lk}}{\bf h}_{jlk}{\underline{{\bf s}}}_{lk}^{\text H}+{\bf N}_{j},\tag{7}
\end{align}
where ${\bf N}_{j}$ represents the $M\times \tau$ additive white Gaussian noise matrix with independent and identically distributed (i.i.d.) zero-mean and unit-variance elements.

Then, in order to estimate the channel between the user $k$ in cell $j$ and the BS in cell $j$, through the LS estimation method \cite{Kay93}, we define
\begin{align}
   \hat{\bf h}_{jjk}^{\text {LS}}\triangleq \frac{1}{\sqrt {{\rho}_{jk}}}{\bf Y}_{j}^{}{\underline{\bf s}}_{jk}\notag,\tag{8}
\end{align}
where $ \hat{\bf h}_{jjk}^{\text {LS}}$ is the estimated  vector of channel ${\bf h}_{jjk}$ based on the LS estimation method.
Therefore, substituting (5)-(7) into (8), we obtain
\begin{align}
     \hat{\bf h}_{jjk}^{\text {LS}}={\bf h}_{jjk}+\sum_{l\neq j}^{L}\frac{\sqrt {{\rho}_{lk}}}{\sqrt {{\rho}_{jk}}}{\bf h}_{jlk}+\frac{1}{\sqrt {{\rho}_{jk}}}{\bf N}_{j}{\underline{{\bf s}}}_{jk}.\tag{9}
\end{align}
Moreover, the MMSE estimation method can be used to estimate ${\bf h}_{jjk}$ with the help of $\hat{\bf h}_{jjk}^{\text {LS}}$ \cite{Kay93}. Hence, we define
\begin{align}
      \hat{\bf h}_{jjk}^{\text {MMSE}}\triangleq\mathbb{E}\left\{{\bf h}_{jjk}(\hat{\bf h}_{jjk}^{\text {LS}})^{\text H}\right\}\left(\mathbb{E}\left\{\hat{\bf h}_{jjk}^{\text {LS}}  (\hat{\bf h}_{jjk}^{\text {LS}})^{\text H}\right\}\right)^{-1}\hat{\bf h}_{jjk}^{\text {LS}}, \tag{10}
\end{align}
to denote the estimator of the channel ${\bf h}_{jjk}$ based on the MMSE estimation method. Then,  it is easy to prove that ${\bf N}_{j}{\underline{{\bf s}}}_{jk}\sim\mathcal {C}\mathcal {N}({\bf 0}_{M\times 1},{\bf I}_{M})$ in (9).  By substituting (3) and (9) into (10),  after some manipulations, we get
\begin{align}
     \hat{\bf h}_{jjk}^{\text {MMSE}}=\frac{\rho_{jk}\beta_{jjk}}{\sum\limits_{l=1}^{L}\rho_{lk}\beta_{jlk}+1}{\hat {\bf h}}_{jjk}^{\text {LS}}. \tag{11}
\end{align}

In the following, we denote the estimator of  ${\bf h}_{jjk}$ as ${\hat{\bf h}}_{jjk}$ for both the LS and MMSE estimation methods, except otherwise denoted.

\subsection{Uplink Data Transmission}
After channel estimation, the BS in each cell uses the obtained CSI to detect the received signal of (1). We consider the standard linear detector maximal-ratio combining (MRC) \cite{Ngo13}. Hence, for the BS in cell $j$, the received
signal ${\bf y}_{j}$ in (1) is separated into $K$ streams by multiplying it with the MRC detector, that is,
\begin{align}
     {\bf r}_{j}=\hat{\bf H}_{jj}^{\text H}{\bf y}_{j}\in\mathbb{C}^{K \times 1},\tag{12}
\end{align}
where
\begin{align}
     \hat{\bf H}_{jj}\triangleq[\hat{\bf h}_{jj1},\ldots,\hat{\bf h}_{jjk} ,\ldots,\hat{\bf h}_{jjK}]\in\mathbb{C}^{M \times K}.\tag{13}
\end{align}
Then, expanding ${\bf y}_{j}$ and denoting the $n$th component value of ${\bf x}_{l}$ as $x_{ln}$, the $k$th entry ${r}_{jk}$ of ${\bf r}_{j}$  can be written as
\begin{align}
     {r}_{jk}&=\underbrace{\sqrt{\rho_{u}}\hat{\bf h}_{jjk}^{\text H}{\bf h}_{jjk}x_{jk}}_{\text {Desired signal}}+\underbrace{\sqrt{\rho_{u}}\sum\limits_{n\neq k}^{K}\hat{\bf h}_{jjk}^{\text H}{\bf h}_{jjn}x_{jn}}_{W_1: \ \text {Intra-cell interference}}+\underbrace{\sqrt{\rho_{u}}\sum\limits_{l\neq j}^{L}\sum\limits_{n=1}^{K}\hat{\bf h}_{jjk}^{\text H}{\bf h}_{jln}x_{ln}}_{W_2: \ \text {Inter-cell interference}}+\underbrace{\hat{\bf h}_{jjk}^{\text H}{\bf n}_{j}}_{W_3: \ \text {Noise}}.\tag{14}
\end{align}

Since we want to investigate the achievable uplink rate of the $k$th user in cell $j$, we assume the term $\mathbb{E}\left\{\hat{\bf h}_{jjk}^{\text H}{\bf h}_{jjk}\right\}$ is perfectly known at the BS $j$. Hence, ${r}_{jk}$ is written as
\begin{align}
     {r}_{jk}=\underbrace{\sqrt{\rho_{u}}\mathbb{E}\left\{\hat{\bf h}_{jjk}^{\text H}{\bf h}_{jjk}\right\}x_{jk}}_{\text {Effective signal}}+\underbrace{\tilde{n}_{jk}}_{\text {Equivalent noise}},\tag{15}
\end{align}
where
\begin{align}
     \tilde{n}_{jk}&\!\!\triangleq\!\!\sqrt{\rho_{u}}\left\{\hat{\bf h}_{jjk}^{\text H}{\bf h}_{jjk}-\mathbb{E}\left\{\hat{\bf h}_{jjk}^{\text H}{\bf h}_{jjk}\right\}\right\}\!x_{jk}\!+\!W_1\!+\!W_2\!+\!W_3.\!\!\tag{16}
\end{align}
From (14)-(16), we can infer that the ``Effective signal" is uncorrelated with the ``Equivalent noise". Hence, using the definition of the effective SINR in multi-cell massive MIMO systems as in \cite[Eq. (12)]{Bjornson15} and considering the effects of pilot overhead, the total
bandwidth, the frequency reuse factors, and the overhead of the cyclic prefix as in \cite[Eq. (14)]{Marzetta10}, the achievable uplink rate of user $k$ in cell $j$, in units of bits/sec, is given by
\begin{align}
    {R}_{jk}=\left(\frac{B}{\Gamma}\right)\left(\frac{T_s-T_p}{T_s}\right)\left(\frac{T_u}{T_o}\right)\log_{2}\left(1+{\text {SINR}}_{jk}\right),\tag{17}
\end{align}
where $\Gamma$ is the frequency reuse factor, $B$ is the total bandwidth, $T_s$ is the  slot length, $T_p$ is the time spent transmitting pilot sequences, $T_u$ is the useful symbol duration, and $T_o$ is the OFDM symbol interval. Also, ${\text {SINR}}_{jk}$ is defined as
\begin{align}
{\text {SINR}}_{jk}\triangleq\frac{\rho_{u}\left|\mathbb{E}\left\{\hat{\bf h}_{jjk}^{\text H}{\bf h}_{jjk}\right\}\right|^{2}}{\rho_{u}\sum\limits_{l=1}^{L}\sum\limits_{n=1}^{K}\mathbb{E}\left\{\left|\hat{\bf h}_{jjk}^{\text H}{\bf h}_{jln}\right|^{2}\right\}-\rho_{u}\left|\mathbb{E}\left\{\hat{\bf h}_{jjk}^{\text H}{\bf h}_{jjk}\right\}\right|^{\!2}+\mathbb{E}\left\{\left\|\hat{\bf h}_{jjk}\right\|_{2}^{2}\right\}}.\tag{18}
\end{align}
The following theorem presents a closed-form expression for ${\text {SINR}}_{jk}$ for both the LS and MMSE estimation methods.
\vskip3mm
{\emph {Theorem 1:}} The exact ${\text {SINR}}_{jk}$, for both the LS and MMSE estimation methods, can be analytically evaluated as
\begin{align}
{\text {SINR}}_{jk}\!=\!\frac{M\rho_{jk}\beta_{jjk}^{2}}{M\!\sum\limits_{l\neq j}^{L}\!\rho_{lk}\beta_{jlk}^{2}\!+\!\!\left(\sum\limits_{l=1}^{L}\!\rho_{lk}\beta_{jlk}\!+\!1\!\right)\!\!\left(\!\frac{1}{\rho_{u}}\!+\!\sum\limits_{l=1}^{L}\!\sum\limits_{n=1}^{K}\!\beta_{jln}\!\right)}.\tag{19}
\end{align}

{\emph {Proof:}} See Appendix A.   \hfill\rule{3mm}{3mm}
\vskip3mm
Note that the exact expression in {\emph {Theorem 1}} can be easily evaluated since it involves only the pilot power, data power, and  large-scale fading coefficients, as well as, $M$. Interestingly, from (19), the achievable uplink rate for both the LS and MMSE estimation methods are identical, since there are linear correlation between the estimators based on the LS and MMSE estimation methods (See (11)) that the numerator and the denominator in (18), for the case of MMSE, can be divided by the same constant $({\rho_{jk}\beta_{jjk}})^2/ ({\sum\limits_{l=1}^{L}\rho_{lk}\beta_{jlk}+1})^2$, which leads the case of MMSE is equal to the case of LS. Also, the deep-rooted reason for (11) is that we consider the Rayleigh fading model with no correlation between the antennas in the BS, which means that the correlated matrix of the user channel is a scaled identity matrix, i.e., $\mathbb{E}\{{\bf h}_{jlk}{\bf h}_{jlk}^{\text H}\}=\beta_{jlk}{\bf I}_{M}$. Besides, the phenomenon of the LS and the MMSE can achieve the same achievable uplink rate for the detector MRC in massive MIMO systems, to the best of our knowledge, was firstly proposed in \cite{Neumann15}. However, the authors in \cite{Neumann15} just gave a  qualitative explanation, while we also give the quantitative analysis in {\emph {Theorem 1}}. Moreover, when the user channel is not like the above mentioned model, which means that the correlated matrix of the user channel is not a scaled identity matrix, the situation of the uplink rate\footnote{In \cite{Yin14}, the authors consider a multi-path model that the angles of arrival are bound. It shows that the MMSE case is better than the LS case for the uplink sum-rate based on the MRC detector.} is outside the scope of this paper. As the next result shows, the exact ${\text {SINR}}_{jk}$  admits further simplifications in the large antennas regime.
\vskip3mm
{\emph {Corollary 1:}} When $M\rightarrow\infty$, the exact analytical expression in (19) approaches to
\begin{align}
\lim_{M\rightarrow\infty}{\text {SINR}}_{jk}=\frac{\rho_{jk}\beta_{jjk}^{2}}{\sum\limits_{l\neq j}^{L}\rho_{lk}\beta_{jlk}^{2}}.\tag{20}
\end{align}

{\emph {Proof:}} The proof is completed by calculating the limit of (19) when $M\rightarrow\infty$. \hfill\rule{3mm}{3mm}
\vskip3mm

It is important to note from (20) that in the high $M$ regime, the PC interference from the other cells' users, whose take the same pilot sequence as the user in the target cell, is the only limit for the achievable uplink rate performance. Moreover, when $\rho_{jk}=\rho_{lk}, \forall j,l$, the ${\text {SINR}}_{jk}$ is equal to the signal-to-interference ratio (SIR) based on LS estimation in \cite[Eq. (13)]{Marzetta10} and the SIR based on MMSE estimation in \cite[Eq. (32)]{Ngo132}, respectively. Hence, based on the above mentioned,  the conclusion of the {\emph {Corollary 1}}  is not only obvious and proper, but also gives a universal formula for ${\text {SINR}}_{jk}$ when $M\rightarrow\infty$ for any pilot power setting.

\section{Pilot Power Allocation}
In this section, we introduce the RCEE metric and, thereafter, aim to find a PPA scheme to minimize the average $\rm {Exp}_{rcee}$ per user. Based on the closed-form expressions for $\rm {Exp}_{rcee}$, we propose a PPA  algorithm to solve this optimization problem.

\subsection{RCEE}
We can now define the RCEE, which is basically a modified NMSE metric, of user $k$ in cell $j$ as follows
\begin{align}
   {\Lambda}_{jk}\triangleq\frac{\|{\bf h}_{jjk}-\hat{{\bf h}}_{jjk}\|_{2}^{2}}{\|{\bf h}_{jjk}\|_{2}^{2}},\tag{21}
\end{align}
and also, its expectation or $\rm {Exp}_{rcee}$ given by
\begin{align}
   {\Delta}_{jk}\triangleq\mathbb{E}\left \{{\Lambda}_{jk}\right\}. \tag{22}
\end{align}
The former metric indicates the instantaneous relative change between the channel estimation error and the true channel in any coherence interval, while the latter metric indicates this relative change over many coherence intervals. The following theorem gives a closed-form expression for $\rm {Exp}_{rcee}$.
\vskip3mm
{\emph {Theorem 2:}} The term ${\Delta}_{jk}$ can be analytically evaluated as
\begin{align}
\begin{split}
    {\Delta}_{jk}\!\!=\!\!\left\{\!\!\!\!
    \begin{array}{*{20}l}
    \infty,\!\!\!\!\!\!\!\!\!\!\!\!&{\text {LS}}\&{\text {MMSE}}, M=1,\\
    \\
   \frac{M\left(\sum\limits_{l\neq j}^{L}{\rho_{lk}}{\beta_{jlk}}+1\right)}{\left(M-1\right){\rho_{jk}}{\beta_{jjk}}}, \!\!\!\!\!\!\!\!\!&{\text {LS}},  M\geq2,\\
    \\
    \frac{\left(\!\!\sum\limits_{l\neq j}^{L}\!\!\rho_{lk}\!\beta_{jlk}\!+\!1\!\!\right)\!\!\left(\!\!\sum\limits_{l\neq j}^{L}\!\!\rho_{lk}\!\beta_{jlk}\!+\!1\!+\!\frac{M\!\rho_{jk}\!\beta_{jjk}}{M-1}\!\!\!\right)}{\left(\sum\limits_{l=1}^{L}\rho_{lk}\beta_{jlk}+1\right)^2}, \!\!\!\!\!& {\text {MMSE}},\!\!\ M\geq2.
     \end{array}\right.
 \end{split}\tag{23}
\end{align}

{\emph {Proof:}} See Appendix B.   \hfill\rule{3mm}{3mm}
\vskip3mm
Note that the formulas in (23) are only meaningful when $M\geq 2$. Moveover, the exact analytical expressions in {\emph {Theorem 2}} can be easily evaluated as they involve only the number of BS antennas,  pilot power, as well as, the large-scaling fading coefficients. Compared with ${\Delta}_{jk}^{\text {LS}}$, ${\Delta}_{jk}^{\text {MMSE}}$ is more complicated.\footnote{Here, for convenience, we use the notation ${\Delta}_{jk}^{\text {LS}}$ and ${\Delta}_{jk}^{\text {MMSE}}$ to replace ${\Delta}_{jk}$ based on LS and MMSE estimation methods, respectively.} Hence, the following corollary gives an upper bound for ${\Delta}_{jk}^{\text {MMSE}}$ in the $M\geq 2$ regime.
\vskip3mm
{\emph {Corollary 2:}} In the $M\geq 2$ regime, ${\Delta}_{jk}^{\text {MMSE}}$ satisfies
\begin{align}
{\Delta}_{jk}^{\text {MMSE}}<\tilde{{\Delta}}_{jk}^{\text {MMSE}}=\frac{M\left(\sum\limits_{l\neq j}^{L}{\rho_{lk}}{\beta_{jlk}}+1\right)}{\left(M-1\right)\left(\sum\limits_{l= 1}^{L}{\rho_{lk}}{\beta_{jlk}}+1\right)}.\tag{24}
\end{align}

{\emph {Proof:}} The proof is trivial and  thus omitted.   \hfill\rule{3mm}{3mm}
\vskip3mm
It is important to note that  $\tilde{{\Delta}}_{jk}^{\text {MMSE}}$ in {\emph {Corollary 2}} will be particularly useful for the PPA problem in subsection III.C.
\vskip3mm
{\emph {Corollary 3:}} When $M\rightarrow\infty$, the exact analytical expressions in (23) approach to
\begin{align}
    \lim_{M\rightarrow\infty}{\Delta}_{jk}=\bar{\Delta}_{jk}=\left\{
\begin{array}{*{20}l}
    \frac{\sum\limits_{l\neq j}^{L}{\rho_{lk}}{\beta_{jlk}}+1}{{\rho_{jk}}{\beta_{jjk}}}, \ {\text {LS}},\\
    \frac{\sum\limits_{l\neq j}^{L}{\rho_{lk}}{\beta_{jlk}}+1}{\sum\limits_{l=1}^{L}{\rho_{lk}}{\beta_{jlk}}+1},\ {\text {MMSE}},
 \end{array}
  \right.\tag{25}
\end{align}
whilst $\tilde{{\Delta}}_{jk}^{\text {MMSE}}$ also approaches to the same limit as ${\Delta}_{jk}^{\text {MMSE}}$.

{\emph {Proof:}} The proof is completed by calculating the limit of (23) and $\tilde{{\Delta}}_{jk}^{\text {MMSE}}$ when $M\rightarrow\infty$. \hfill\rule{3mm}{3mm}
\vskip3mm
It is interesting to note from  {\emph {Corollary 3}} that as $M$ grows, the $\rm {Exp}_{rcee}$ will decrease and approach a constant value for both the LS and MMSE estimation methods.  Moreover, from (23) and (25), we see that MMSE performs better than LS since the MMSE estimation method utilizes the additional second-order statistical information.
\vskip3mm
{\emph {Corollary 4:}} When $M\rightarrow\infty$, the relationship between the RCEE and the $\rm {Exp}_{rcee}$ is given by
\begin{align}
{\Lambda}_{jk}\xrightarrow[]{a.s.}\bar{\Delta}_{jk},\tag{26}
\end{align}
where $\xrightarrow[]{a.s.}$ denotes almost sure convergence.

{\emph {Proof:}} The proof follows trivially by utilizing the law of large numbers \cite {Ngo13} and Appendix B.

\hfill\rule{3mm}{3mm}
\vskip3mm
It is important to note from {\emph {Corollary 4}} that,  the RCEE converges to its statistical value ($\rm {Exp}_{rcee}$) when $M$ is large. In other words, the  stochastic nature
of the RCEE disappears when $M$ is big enough. Note that (26) reflects the channel hardening effect via the channel estimation performance, though in a slightly different way than in \cite{Narasimhan14}.

\subsection{Constrained Optimization Problem}
Now, to avoid an iterative non-stationary optimization problem, we focus on the PPA in one cell, while the pilot  power in other cells is kept fixed, and we call it as the PPA scheme. It is clear that the RCEE which is obtained from one time channel estimation is not able to represent the channel estimation performance in a period of time. For the massive MIMO setup consideration, we can avail of the channel hardening effect and work exclusively with the $\rm {{Exp}}_{{rcee}}$ instead of RCEE. Moreover, considering the fairness of all users in the target cell, we choose the average $\rm {{Exp}}_{{rcee}}$ per user as our objective function to evaluate the system channel estimation performance. We can now formulate the following constrained PPA optimization problem
\begin{align}
\begin{split}
   {\text {minimize}} \ \ \ &\frac{1}{K}\sum_{k=1}^{K}{ \Delta}_{jk},\\
   {\text {subject to}}\ \ \ &\sum_{k=1}^{K}{\rho}_{jk}\leq {P},\\
   &{{\rho}_{\rm min}}\leq{\rho}_{jk}\leq{{\rho}_{\rm max}},\forall k=1,\ldots,K,
\end{split}\tag{27}
\end{align}
where $P$ is the total pilot power budget of the users in each cell, ${{\rho}_{\rm min}}$ and ${{\rho}_{\rm max}}$ denote the lower and upper bounds of the variation range of ${\rho}_{jk}$.{\footnote{Here, we assume  ${\rho_{\rm min}}={P}/2K$,  ${\rho_{\rm max}}=\mu{\text P}/K$, and $\mu={\rho_{\rm max}}K/{ P}\in[3/2,(K+1)/2]$.
The term ${P}/2K$ is obtained from the hypothesis that the user's pilot power cannot be below the half of the  average per user pilot power.
The term $(K+1)/2$ is obtained from the hypothesis that $K\geq2$ (i.e. there are at least 2 users in each cell) and $(K-1){\rho_{\rm min}}+{\rho_{\rm max}}\leq {P}$ (It is possible that at least one user in cell can be allocated the maximum pilot power).
The term $3/2$ is obtained from the hypothesis that if one user's pilot power is allocated the maximum pilot power, its pilot power can achieve at least $3/2$ of the average per-user pilot power.}} Once the problem (27) is solved, it means that the pilot power $\rho_{jk}$ is determined.

We now turn our attention to the special case, EPPA scheme, where all cell users have the same pilot power setting ($\rho_{jk}=P/K,\forall j,k$), and obtain the following corollary.
\vskip3mm
{\emph {Corollary 5:}} When  $M\rightarrow\infty$ and for the EPPA scheme, the exact analytical expressions in (23) approach to
\begin{align}
    \lim_{M\rightarrow\infty}{\Delta}_{jk}=\bar{\Delta}_{jk}=\left\{
\begin{array}{*{20}l}
    \frac{\sum\limits_{l\neq j}^{L}{\beta_{jlk}}+\frac{K}{P}}{{\beta_{jjk}}}, \ {\text {LS}},\\
    \frac{\sum\limits_{l\neq j}^{L}{\beta_{jlk}}+\frac{K}{P}}{\sum\limits_{l=1}^{L}{\beta_{jlk}}+\frac{K}{P}},\ {\text {MMSE}}.
 \end{array}
  \right.\tag{28}
\end{align}

{\emph {Proof:}} The proof is completed by calculating the limit of (23) when considering the EPPA scheme and $M\rightarrow\infty$. \hfill\rule{3mm}{3mm}
\vskip3mm

\subsection{Approximate  Unconstrained Solution}
According to {\emph {Theorem 2}}, we have already obtained the asymptotic analytical expression for the objective function in (27). However, ${\Delta}_{jk}^{\text {MMSE}}$ is complicated and the second constraint in (27) makes it challenging to solve. Hence, we now replace ${\Delta}_{jk}^{\text {MMSE}}$ with $\tilde{\Delta}_{jk}^{\text {MMSE}}$ and release the constraint ${{\rho}_{\rm min}}\leq{\rho}_{jk}\leq{{\rho}_{\rm max}}$ in (27). The optimization problem can be rewritten as
\begin{align}
\begin{split}
   {\text {minimize}} \ \ \ &\frac{1}{K}\sum_{k=1}^{K}\tilde{{\Delta}}_{jk},
   \\
   {\text {subject to}}\ \ \ &\sum_{k=1}^{K}{\rho}_{jk}\leq{P},
\end{split}\tag{29}
\end{align}
where
\begin{align}
\begin{split}
    {\tilde{\Delta}}_{jk}\!\triangleq\!\left\{\!
    \begin{array}{*{20}l}
   \frac{M\left(\sum\limits_{l\neq j}^{L}{\rho_{lk}}{\beta_{jlk}}+1\right)}{\left(M-1\right){\rho_{jk}}{\beta_{jjk}}}, &{\text {LS}},\\
    \\
    \frac{M\left(\sum\limits_{l\neq j}^{L}{\rho_{lk}}{\beta_{jlk}}+1\right)}{\left(M-1\right)\left(\sum\limits_{l= 1}^{L}{\rho_{lk}}{\beta_{jlk}}+1\right)}, &{\text {MMSE}}.
     \end{array}\right.
 \end{split}\tag{30}
\end{align}
The following theorem presents closed-form expressions for the optimal solution of (29).
\vskip3mm
{\emph {Theorem 3:}} The optimal solution ${\rho_{jk}^{\ast}}$ of (29) is formulated as
\begin{align}
    {\rho_{jk}^{\ast}}\!=\!\left\{
\begin{array}{*{20}l}
    \frac{1}{\lambda_{\text {LS}}} \left({\frac{\upsilon_{jk}}{\beta_{jjk}}}\right)^{\frac{1}{2}}, \ &{\text {LS}},\\
    \frac{1}{\lambda_{\text {MMSE}}}\left(\frac{\upsilon_{jk}}{\beta_{jjk}}
    \right)^{\frac{1}{2}}- \frac{\upsilon_{jk}}{\beta_{jjk}},&{\text {MMSE}},
 \end{array}
  \right.\tag{31}
\end{align}
where
\begin{align}
   \upsilon_{jk}&\triangleq\sum\limits_{l\neq j}^{L}{\rho_{lk}}{\beta_{jlk}}+1,\tag{32}\\
   \lambda_{\text {LS}}&\triangleq\frac{1}{P}\sum\limits_{k=1}^{K}\left ({\frac{\upsilon_{jk}}{\beta_{jjk}}}\right)^{\frac{1}{2}},\tag{33}\\
   \lambda_{\text {MMSE}}&\triangleq\frac{\sum\limits_{k=1}^{K} \left({\frac{\upsilon_{jk}}{\beta_{jjk}}}\right)^{\frac{1}{2}}}
   {{P}+\sum\limits_{k=1}^{K}\frac{\upsilon_{jk}}{\beta_{jjk}}}.\tag{34}
\end{align}

{\emph {Proof:}} The proof is completed by using  the Lagrange multiplier method \cite{Boyd04}.  \hfill\rule{3mm}{3mm}
\vskip3mm
Note that from (31)-(34), we see that the optimal solution of (29) is independent of the number of BS antennas. Most importantly, the optimal solution in (29) can serve as a very effective starting point for solving the constrained optimization problem (27) approximately. This is one of the main ideas behind the PPA algorithm. The second idea is to partition the users in cell $j$ into three groups. Users in group 1 have the same pilot power $\rho_{\rm min}$. Users in group 2 have the same pilot power $\rho_{\rm max}$. Group 3 users' pilot power is determined by solving (29) with the remaining available total pilot power budget. Inspired by this, we propose a PPA algorithm ({\bf {Algorithm 1}}), based on a simple decision criterion for both the LS and MMSE estimation methods, to obtain an approximate solution of optimization problem (27).
\begin{algorithm}[!t]
\caption{PPA algorithm}
\begin{algorithmic}[1]
\Procedure {PPA}\\
\ \ \ {\bf {input}}: ${P}$, $K$, $\beta_{jlk}$, $\rho_{lk}$ $(l\neq j)$, $\rho_{\rm min}$, $\rho_{\rm max}$, and Flag (${\text {Flag}}=0$ or ${\text {Flag}}=1$).\vspace{-0.6em}\\
\ \ \ {\bf {initialization}}: ${\mathcal {K}}_{{P}}=\{1,\ldots,K\}$, ${\mathcal {K}}_{{P},{\rm min}}=\O$, and ${\mathcal {K}}_{{ P},{\rm max}}=\O$.\vspace{-0.6em}\\
\ \ \ {\bf {if}} ${\text{Flag}}==0$\vspace{-0.6em}\\
\ \ \ \ \ \ Calculate  the first expression in (31).\vspace{-0.6em}\\
\ \ \ {\bf {else}} \vspace{-0.6em}\\
\ \ \ \ \ \ Calculate  the second expression in (31).\vspace{-0.6em}\\
\ \ \ {\bf {end if}}\vspace{-0.6em}\\
\ \ \ {\bf {for}} $i\gets 1, ..., K$ {\bf {do}}\vspace{-0.6em}\\
\ \ \ \ \ \ {\bf {if}} ${\rho_{jk}}\geq{\rho_{\rm min}}$ \& ${\rho_{jk}}\leq{\rho_{\rm max}}$, $\forall k\in {\mathcal {K}}_{P}$\vspace{-0.6em}\\
\ \ \ \ \ \ \ \ \ {\bf {break}}\vspace{-0.6em}\\
\ \ \ \ \ \ {\bf {end if}}\vspace{-0.6em}\\
\ \ \ \ \ \  Obtain $\{k_{i}|\ {\rho_{jk_{i}}}<{\rho_{\rm min}}, k_{i}\in {\mathcal {K}}_{P}\}$, and $\{t_{i}|\ {\rho_{jt_{i}}}>{\rho_{\rm min}}, t_{i}\in {\mathcal {K}}_{P}\}$.\vspace{-0.6em}\\
\ \ \ \ \ \  Obtain $k_{i}^{\star}=\operatorname* {arg \ max}\limits_{k_{i}^{\star}\in\{k_{i}\}}|{\rho_{jk_{i}}}-{\rho_{\rm min}}|$, and $t_{i}^{\star}=\operatorname* {arg \ max}\limits_{t_{i}^{\star}\in\{t_{i}\}}|{\rho_{jt_{i}}}-{\rho_{\rm max}}|$.\vspace{-0.1em}\\
\ \ \ \ \ \ {\bf {if}} (${k_{i}^{\star}}$ \!\&\! ${t_{i}^{\star}}$ \!\&\! $|{\rho_{jk_{i}^{\star}}}-{\rho_{\rm min}}|\geq|{\rho_{jt_{i}^{\star}}}-{\rho_{\rm max}}|$)\ $||$\ $\{t_{i}\}\!=\!\O$\vspace{-0.6em}\\
\ \ \ \ \ \ \ \ \ ${\rho_{jk_{i}^{\star}}}\gets{\rho_{\rm min}}$, $P\gets P-{\rho_{jk_{i}^{\star}}}$, ${\mathcal {K}}_{P}\gets{\mathcal {K}}_{P}\setminus\{k_{i}^{\star}\}$, and ${\mathcal {K}}_{{P},{\rm min}}\gets{\mathcal {K}}_{{P},{\rm min}}\cup\{k_{i}^{\star}\}$.\vspace{-0.6em}\\
\ \ \ \ \ \ \ \ \ {\bf {if}} ${\text{Flag}}==0$\vspace{-0.2em}\\
\ \ \ \ \ \ \ \ \ \ \ \ {{Replace}} $\sum\limits_{k=1}^{K}$ to $\sum\limits_{k\in {\mathcal {K}}_{ P}}$ in (33) and calculate (33). \vspace{-0.1em}\\
\ \ \ \ \ \ \ \ \ \ \ \ {\bf {for}} $k\in{\mathcal {K}}_{P}$ {\bf {do}}\vspace{-0.4em}\\
\ \ \ \ \ \ \ \ \ \ \ \ \ \ \  Calculate  the first expression in (31).\vspace{-0.4em}\\
\ \ \ \ \ \ \ \ \ \ \ \ {\bf {end for}}\vspace{-0.6em}\\
\ \ \ \ \ \ \ \ \ {\bf {else}} \vspace{-0.6em}\\
\ \ \ \ \ \ \ \ \ \ \ \ {{Replace}} $\sum\limits_{k=1}^{K}$ to $\sum\limits_{k\in {\mathcal {K}}_{ P}}$ in (34) and calculate (34).\vspace{-0.1em}\\
\ \ \ \ \ \ \ \ \ \ \ \ {\bf {for}} $k\in{\mathcal {K}}_{P}$ {\bf {do}}\vspace{-0.6em}\\
\ \ \ \ \ \ \ \ \ \ \ \ \ \ \  Calculate  the second expression in (31).\vspace{-0.6em}\\
\ \ \ \ \ \ \ \ \ \ \ \ {\bf {end for}}\vspace{-0.6em}\\
\ \ \ \ \ \ \ \ \ {\bf {end if}}\vspace{-0.6em}\\
\ \ \ \ \ \ {\bf {else }} \vspace{-0.6em}\\
\ \ \ \ \ \ \ \ \ ${\rho_{jt_{i}^{\star}}}\gets{\rho_{\rm max}}$, $P\gets P-{\rho_{jt_{i}^{\star}}}$, ${\mathcal {K}}_{P}\gets{\mathcal {K}}_{P}\setminus\{t_{i}^{\star}\}$, and ${\mathcal {K}}_{{P},{\rm max}}\gets{\mathcal {K}}_{{P},{\rm max}}\cup\{t_{i}^{\star}\}$.\vspace{-0.6em}\\
\ \ \ \ \ \ \ \ \ Execute the procedure of lines 17-27.\vspace{-0.6em}\\
\ \ \ \ \ \ {\bf {end if}}\vspace{-0.6em}\\
\ \ \ {\bf {end for}}\vspace{-0.6em}\\
\ \ \ {\bf {output}}: $\rho_{jk}$, ${\mathcal {K}}_{{P}}$, ${\mathcal {K}}_{{P},{\rm min}}$, and ${\mathcal {K}}_{{P},{\rm max}}$.\vspace{-0.6em}\\
\bf {end procedure}
\end{algorithmic}
\end{algorithm}

In {\bf {Algorithm 1}}, ${\mathcal {K}}_{{P},{\rm min}}$ and ${\mathcal {K}}_{{P},{\rm max}}$ denote the user groups where users' pilot powers are  the minimum pilot power $\rho_{\rm min}$ and the maximum pilot power $\rho_{\rm max}$, respectively. Also, after removing the users in ${\mathcal {K}}_{{P},{\rm min}}$ or ${\mathcal {K}}_{{P},{\rm max}}$, the remaining users belong to the user group ${\mathcal {K}}_{{P}}$. Users' pilot powers have the following relationship
\begin{align}
\sum\limits_{\mathclap{{k\in {\mathcal {K}}_{{P},{\rm min}}}}}{\rho_{\rm min}}+\sum\limits_{\mathclap{k\in {\mathcal {K}}_{{P},{\rm max}}}}{\rho_{\rm max}}+\sum\limits_{\mathclap{{k\in {\mathcal {K}}_{{ P}}}}}{\rho_{jk}}={P}.\tag{35}
\end{align}
These groups are initialized in {\bf {Algorithm 1}}.
The cardinalities of these groups are denoted as $K_{{P},\rm min}$, $K_{{P},\rm max}$ and $K_{P}$, respectively, which satisfy
\begin{align}
K_{P}+K_{{P},\rm min}+K_{{P},\rm max}=K.\tag{36}
\end{align}
Generally speaking, the core idea of the PPA algorithm is to give an approximately optimal partition of the users in the target cell through the quality of their channels.
\begin{figure}[!t]
\centering
\includegraphics[width=3.5in]{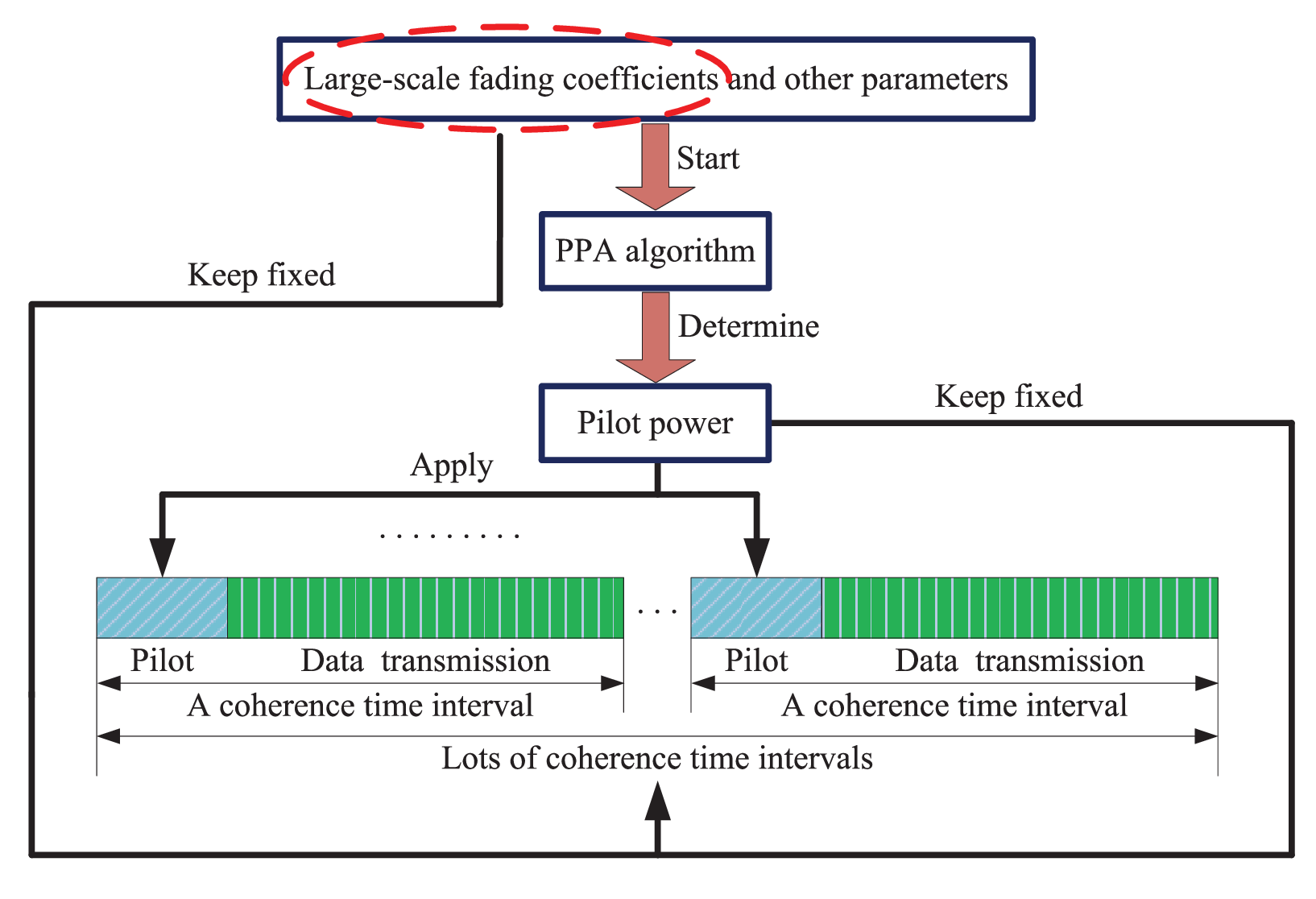}
\caption{The procedure of the PPA algorithm.}
\label{dropped}
\end{figure}

Note that the entire procedure of the PPA algorithm is depicted in Fig. 1. Given the fact that in the massive MIMO regime, the stochastic nature of RCEE vanishes, large-scale fading coefficients are the main parameters in our objective function in (27). This observation simplifies the problem of PPA substantially, since we can simply put the large-scale fading coefficients and other parameters into the PPA algorithm and determine the pilot power of the users in the target cell. Moreover, we have considered the block fading assumption such that the large-scale fading coefficients remain fixed over lots of coherence time intervals. Hence, once we start the PPA algorithm and obtain the pilot power, regardless of the small-scale variations between two consecutive coherence intervals, our pilot power keeps fixed within these coherence time intervals.

Since we propose a PPA  algorithm to allocate pilot power in the target cell to minimize the average $\rm {{Exp}}_{{rcee}}$ per user for a given $P$, the  limiting performance will also be obtained with infinitely high $P$. Hence, the following theorem gives the asymptotic expression for $\rm {Exp}_{rcee}$ when both $P$ and $M$ tend to infinity.\footnote{Note that the relationship between the RCEE and  $\rm {{Exp}}_{{rcee}}$ has already been given in {\emph {Corollary 4}} when $M\rightarrow\infty$. Hence, for convenience, we just study the performance of $\rm {{Exp}}_{{rcee}}$ in the subsequent parts.}
\vskip3mm
{\emph {Theorem 4:}} When $M\rightarrow\infty$ and $P\rightarrow\infty$, and with the help of (25), $\rm {Exp}_{rcee}$ is asymptotically approximated by
\begin{align}
\lim_{P\rightarrow\infty}\bar\Delta_{jk}=\breve{{\Delta}}_{jk}^{\text {LS}}\approx
\left\{
{
\begin{array}{*{20}l}
    \frac{\sum\limits_{l\neq j}^{L}{\delta_{lk}}{\beta_{jlk}}}{\alpha\beta_{jjk}}, \  &k\in\dot{\mathcal {K}}_{\rm min},\\
    \frac{\sum\limits_{l\neq j}^{L}{\delta_{lk}}{\beta_{jlk}}}{\mu\beta_{jjk}}, \ &k\in\dot{\mathcal {K}}_{\rm max},\\
    \frac{\phi}{\varphi\psi}, \ &k\in\dot{\mathcal {K}}_{},
 \end{array}
 } \right.\tag{37}
\end{align}
with LS estimation, and
\begin{align}
\lim_{P\rightarrow\infty}\bar\Delta_{jk}=\breve{{\Delta}}_{jk}^{\text {MMSE}}\approx
\left\{
{
\begin{array}{*{20}l}
    \frac{\sum\limits_{l\neq j}^{L}{\delta_{lk}}{\beta_{jlk}}}{\sum\limits_{l\neq j}^{L}{\delta_{lk}}{\beta_{jlk}}+\alpha\beta_{jjk}}, \  &k\in\dot{\mathcal {K}}_{\rm min},\\
    \frac{\sum\limits_{l\neq j}^{L}{\delta_{lk}}{\beta_{jlk}}}{\sum\limits_{l\neq j}^{L}{\delta_{lk}}{\beta_{jlk}}+\mu\beta_{jjk}}, \ &k\in\dot{\mathcal {K}}_{\rm max},\\
    \frac{\phi}{\left(\varphi+\varpi\right)\psi}, \ &k\in\dot{\mathcal {K}}_{},
 \end{array}
 } \right.\tag{38}
\end{align}
with MMSE estimation, where $\delta_{lk}\in (0,1)$ is the scale factor which satisfies $\delta_{lk}={\rho_{lk}}/{P}$ for $l\neq j$, $\alpha={\rho_{\rm min}}K/{P}=0.5$, and
\begin{align}
    \phi&\triangleq\sum\limits_{l\neq j}^{L}{\delta_{lk}}{\beta_{jlk}}\sum\limits_{k\in \dot{\mathcal {K}}_{}}{{ \left({\frac{1}{\beta_{jjk}}{\sum\limits_{l\neq j}^{L}{\delta_{lk}}{\beta_{jlk}}}}\right)^{\frac{1}{2}}}},\tag{39}\\
    \varphi&\triangleq1-\frac{{\alpha}\dot{K}_{\rm min}+{\mu}\dot{K}_{\rm max}}{K}
    ,\tag{40}\\
    \psi&\triangleq\left({\beta_{jjk}\sum\limits_{l\neq j}^{L}{\delta_{lk}}{\beta_{jlk}}}\right)^{\frac{1}{2}},\tag{41}\\
    \varpi&\triangleq\sum\limits_{k\in \dot{\mathcal K}_{}}\frac{\sum\limits_{l\neq j}^{L}{\delta_{lk}}{\beta_{jlk}}}{\beta_{jjk}}.\tag{42}
\end{align}
Also, $\dot{\mathcal {K}}_{\rm min}$, $\dot{\mathcal {K}}_{\rm max}$, $\dot{\mathcal {K}}_{}$, $\dot{ {K}}_{\rm min}$, $\dot{{K}}_{\rm max}$, and $\dot{ {K}}_{}$ are defined in Appendix D.\footnote{In fact,  the user groups for the LS and MMSE estimation methods may be different. However, for convenience, we still use these groups $\dot{\mathcal {K}}_{\rm min}$, $\dot{\mathcal {K}}_{\rm max}$, and $\dot{\mathcal {K}}_{}$ to represent both the LS and MMSE estimation methods.}

{\emph {Proof:}} See Appendix C. \hfill\rule{3mm}{3mm}
\vskip3mm
It is important to note that, as $M\rightarrow\infty$ and $P\rightarrow\infty$,   $\rm{Exp}_{rcee}$ is approximately controlled by the   large-scale fading coefficients  and the ratio of the user's pilot  power to the total pilot power budget. Also, by comparing (37) and (38), for the user groups $\dot{\mathcal {K}}_{\rm min}$ and $\dot{\mathcal {K}}_{\rm max}$, respectively, the  denominator  of (38) contains the impact of the  users'  large-scale fading coefficients in the other cells compared with the  denominator  of (37). The main reason of this phenomenon is that the MMSE estimation method accounts for inter-cell interference (the large-scale fading from other cells), whereas the LS estimation method treats the interference from other cells as noise. Interestingly, for the  group $\dot{\mathcal {K}}_{}$, the main difference between the third expression for (37) and (38) is the term $\varpi$ since the MMSE estimation method considers the effect of $\beta_{jlk}$ ($l\neq j$).

We now consider the special case of the EPPA scheme. The following corollary evaluates RCEE and $\rm {Exp}_{rcee}$ with EPPA scheme, when $M$ and $P$ are asymptotically large.
\vskip3mm
{\emph {Corollary 6:}} When $M\rightarrow\infty$ and $P\rightarrow\infty$, for the EPPA scheme,
\begin{align}
\lim_{P\rightarrow\infty}\bar\Delta_{jk}=
\left\{
{
\begin{array}{*{20}l}
\frac{\sum\limits_{l\neq j}^{L}{\beta_{jlk}}}{{\beta_{jjk}}},\ {\text {LS}},\\
\frac{\sum\limits_{l\neq j}^{L}{\beta_{jlk}}}{\sum\limits_{l=1}^{L}{\beta_{jlk}}}, \ {\text {MMSE}}.
 \end{array}
 } \right.\tag{43}
\end{align}

{\emph {Proof:}} The proof is completed by calculating the limit of (28) when $P\rightarrow\infty$. \hfill\rule{3mm}{3mm}
\vskip3mm
Note that in the EPPA scheme, the expressions for $\rm {Exp}_{rcee}$ with infinite $M$ and $P$ are simpler than that in PPA scheme. Moreover, (43) provides physical insights into the channel estimation performance for the LS and MMSE estimation methods, respectively, to reflect the PC effect; interestingly, these expressions are inversely analogous to the SIR in \cite[Eq. (13)]{Marzetta10}.

\section{Numerical Results}
In this section, we consider a hexagonal cellular network  with a set of $L$ cells and radius $r$ meters where users are distributed uniformly in each cell. Also, the large-scale fading coefficients, which account for geometric attenuation and shadow fading, are set as \cite{Ngo141}
\begin{align}
   \beta_{jlk}=\frac{z_{jlk}}{1+(r_{jlk}/r_{\rm min})^{\gamma}},\tag{44}
\end{align}
where $z_{jlk}$ is a log-normal random variable with standard deviation $\sigma$, $\gamma$ is the path loss exponent, $r_{jlk}$ is the distance between the user $k$ in cell $l$ and the BS in cell $j$, and $r_{\rm min}$ is the reference distance. In our simulations, we choose $L=7$, $r=500$m, $\sigma=8$dB, $\gamma=3.8$, and $r_{\rm min}=200$m, which also follow the methodology of \cite{Ngo141}.

\subsection{Verification of the PPA Algorithm}
\begin{figure}[!t]
\centering
\includegraphics[width=3.6in]{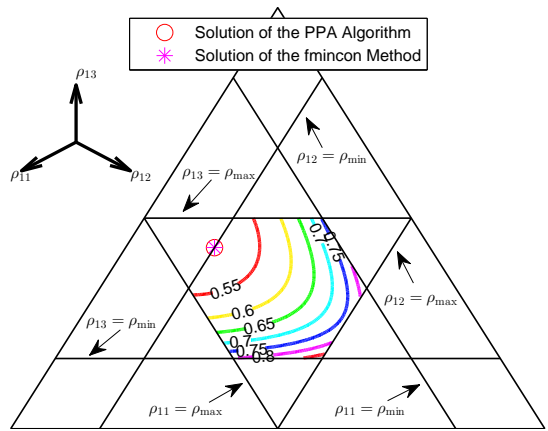}
\caption{The objective function  values in (27) and the solutions of PPA algorithm and fmincon method are shown in the form of contour lines.}
\label{dropped}
\end{figure}

\begin{table*}
\caption{}
\centering
\subtable[List of the large-scale coefficients]{
       \begin{tabular}{|c|c|c|c|}
\hline
\diagbox{Cell $l$}{$\beta_{1lk}$}{User $k$} & 1 & 2 & 3\\
\hline
1   & 0.0304  & 1.2899   & 0.0655\\
\hline
2   & 0.0006  & 0.0290   & 0.0389\\
\hline
3   & 0.0045  & 0.0024   & 0.0070\\
\hline
4   & 0.0080  & 0.0039   & 0.0045\\
\hline
5   & 0.0008  & 0.0842   & 0.0028\\
\hline
6   & 0.0078  & 0.0003   & 0.0026\\
\hline
7   & 0.0011  & 0.0177   & 0.0014\\
\hline
\end{tabular}
}
\qquad
\subtable[List of the average runtime values for solving (27)]{
       \begin{tabular}{|c|c|c|}
\hline
\multicolumn{1}{|c|}{}& \multicolumn{2}{|c|}{{Average runtime$\times10^{-5}$(s)}}\\
\hline
{\diagbox{$K$}{Method}} & {\text {PPA}} & {\text {fmincon}}\\
\hline
2  &2.2    & 1239.2     \\
\hline
3  &4.0    & 1354.6     \\
\hline
4  &6.4    & 1405.4     \\
\hline
5  &9.1    & 1471.3     \\
\hline
6  &12.3   & 1573.4     \\
\hline
7  &16.2   & 1702.1     \\
\hline
8  &20.8   & 1898.0     \\
\hline
9  &25.6   & 1966.2     \\
\hline
10 &31.4   & 2018.7     \\
\hline
\end{tabular}
}
\end{table*}

In this subsection, we will check the PPA algorithm for the LS method based on {\emph {Theorem 3}},  while the evaluation based on the MMSE estimation method can be omitted due to similarity.  In order to give a visual display of our PPA algorithm to solve (27), the number of users $K$ in each cell is assumed to be 3. Since we assume that the noise variance is 1, $P$ can be interpreted as the transmit signal to noise ratio (SNR) and, thus, can be expressed in dB. For convenience, we set the average per user pilot power ${P}/K$ to 30 dB. Also, we set the pilot sequence length to $\tau=K$, $\mu=1.5$, and $M=200$.

We choose the center cell of the 7-cell hexagonal cellular network as our objective and call it cell 1, which means we choose $j=1$. The large-scale fading coefficients of the channel between the users in  cell 1-7 and the BS in cell 1 are randomly generated and given in Table Ia.

We use the PPA algorithm to allocate the three users' pilot power in cell 1. In order to validate the effectiveness of the PPA algorithm, we also compute the approximate solution of the constrained optimization problem in (27) through the MATLAB function ``fmincon". Note that we use the solution of (29) as the start point for fmincon.
Based on the monotonicity of the objective function and the constraints in (27), it is obvious that the global optimal solution point of (27) is at the hyperplane
\begin{align}
\sum\limits_{k=1}^{K}{\rho_{1k}}=P,\  \rho_{\rm min}\leq\rho_{1k}\leq\rho_{\rm max}, \ \forall k=1,\ldots,K.\tag{45}
\end{align}
Therefore, considering that the objective function in (27) now has three variables, we use the method of projection in the form of contour lines  to show the value of the objective function in (27) in  the plane $\rho_{11}+\rho_{12}+\rho_{13}=P$.

Fig. 2 gives the objective function  values in (27) in the form of contour lines. Also, the solution points of the PPA algorithm and fmincon method  are marked by a red circle and a black asterisk, respectively. This graph shows that the geometrical centers of these two mark symbols overlap almost completely, which implies that the solution obtained from the PPA algorithm is almost the same as the one obtained from the numerically evaluated method.

We now turn our attention to the issue of computational complexity by computing the average complexity of both techniques.  The average runtime of each algorithm is obtained by averaging over 1000 independent large-scale coefficient realizations. The average runtime values for both PPA algorithm and fmincon method, as well as, different number of users in cell 1, are presented in Table Ib. We can see that, when the number of users $K$ gets larger,  the average runtime increases for all cases under consideration. Most importantly, compared with the fmincon method, the PPA algorithm's average runtime is significantly reduced, (e.g., compared with the fmincon method, the average runtime of the proposed PPA algorithm decreases about 600 times for $K=2$ and 70 times for $K=10$). Hence, our PPA algorithm offers the important advantage of low computational complexity.

\subsection{Channel Estimation Performance  Comparison}
\begin{figure*}[!t]
\subfigure[LS estimation method] { \label{fig:(a)}
\includegraphics[width=3.1in,height=2.1in]{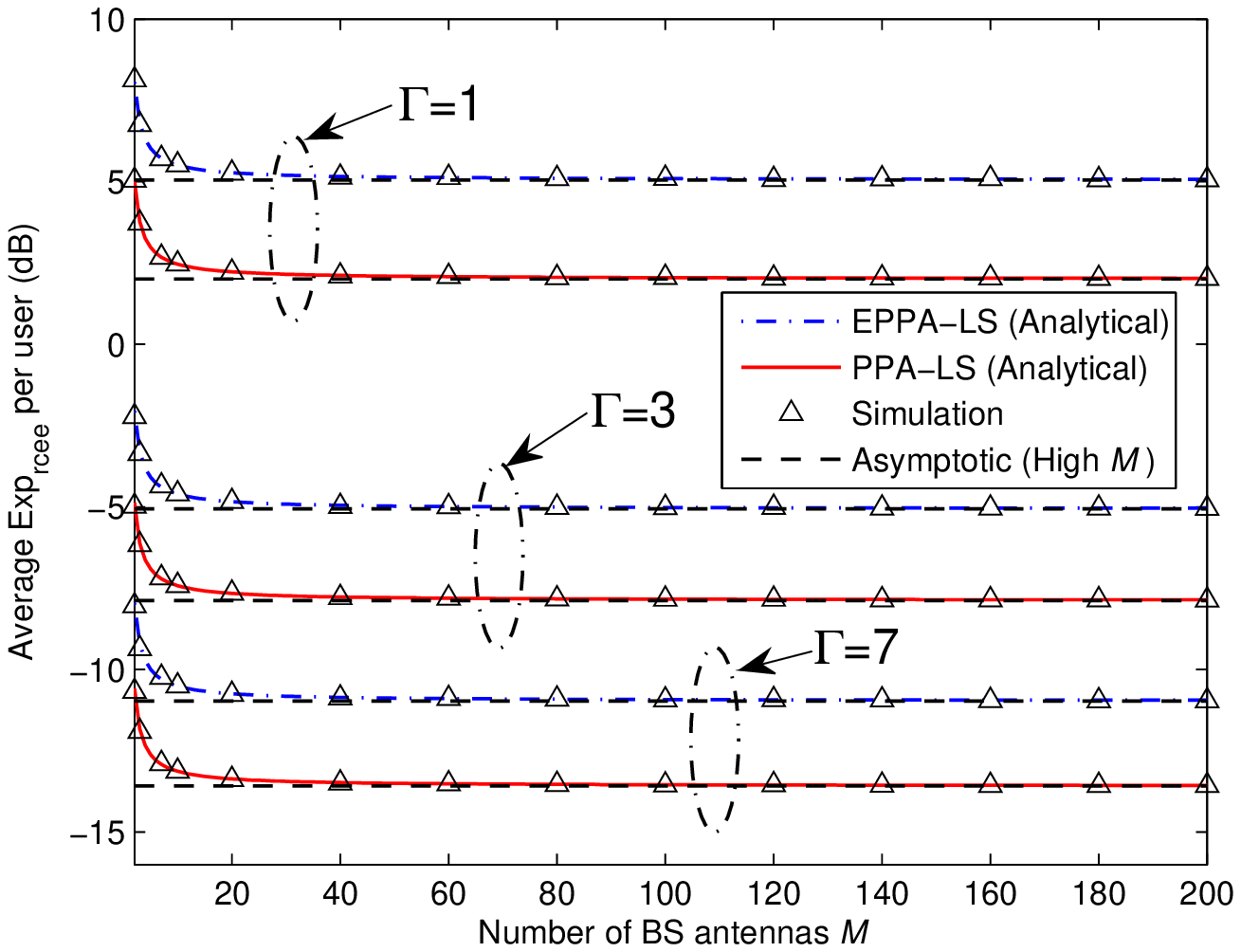}
}
\subfigure[MMSE estimation method] { \label{fig:(b)}
\includegraphics[width=3.1in,height=2.1in]{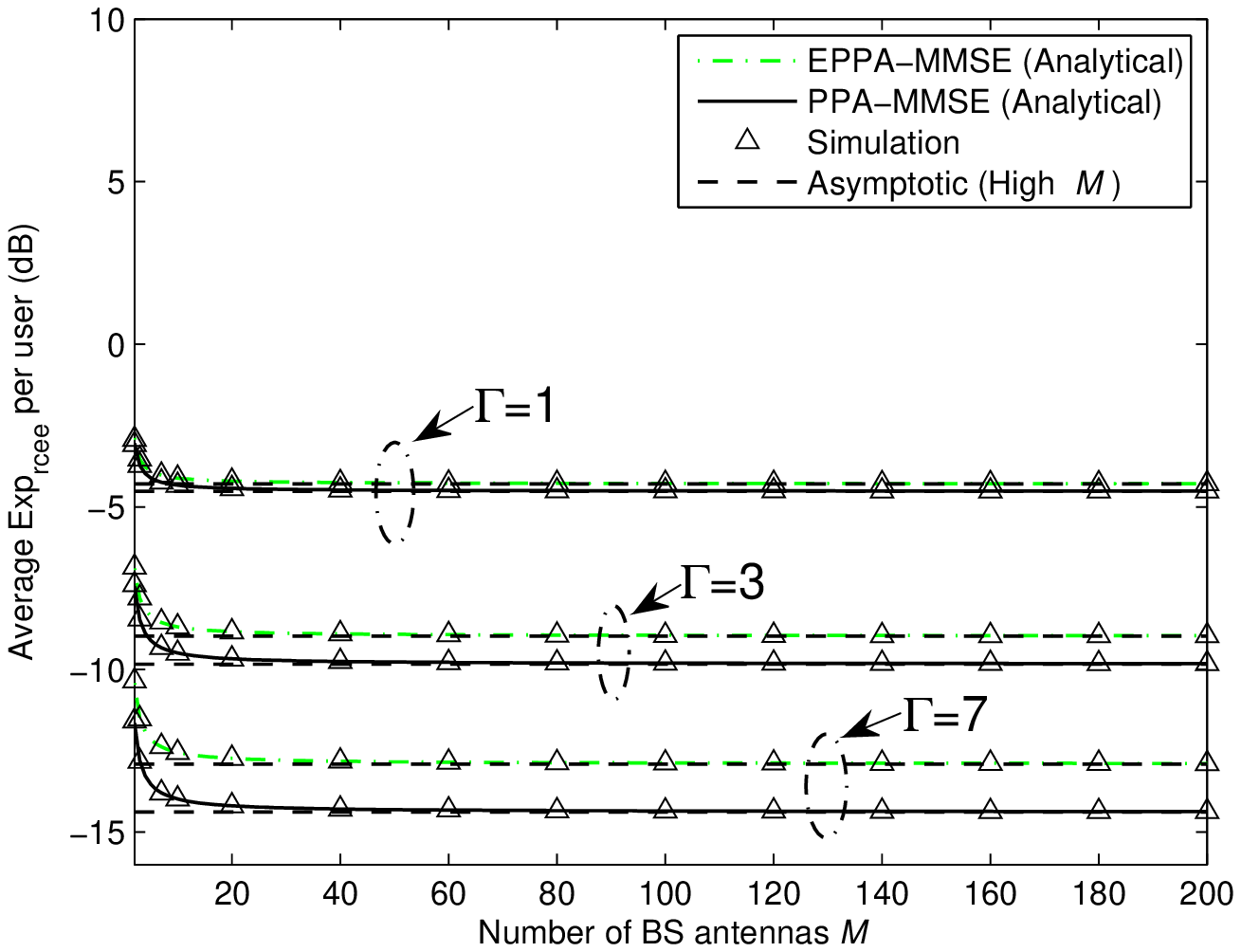}
}
\caption{The average $\rm { {Exp}}_{ {rcee}}$ per user as a function of the number of BS antennas with $P=40$ dB for both the EPPA and PPA schemes.}
\label{fig}
\end{figure*}

\begin{figure*}[!t]
\subfigure[CDF of average $\rm {{Exp}}_{ {rcee}}$ per user] { \label{fig:(a)}
\includegraphics[width=3.1in,height=2.1in]{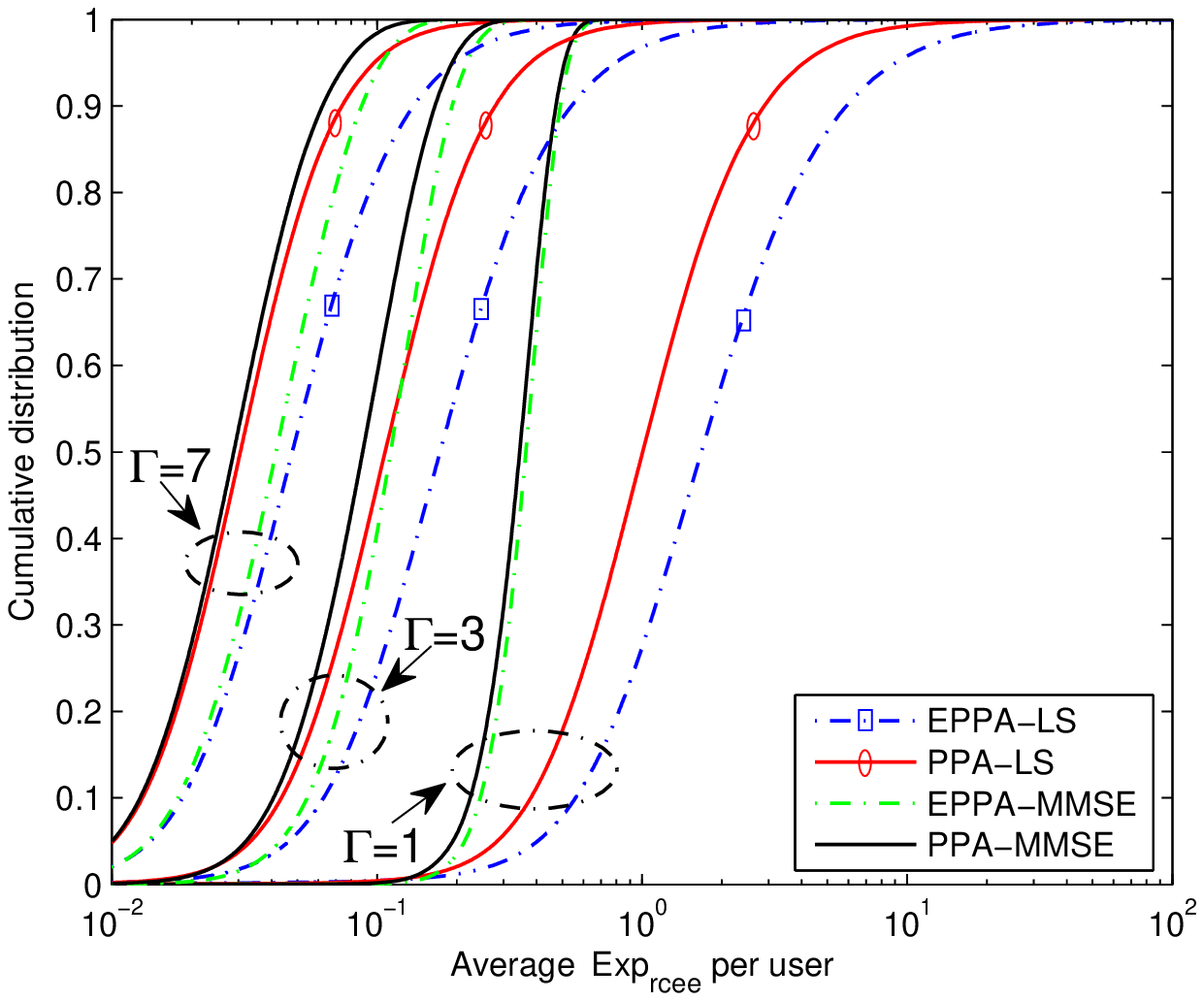}
}
\subfigure[Average $\rm {{Exp}}_{{rcee}}$ per user] { \label{fig:(b)}
\includegraphics[width=3.1in,height=2.1in]{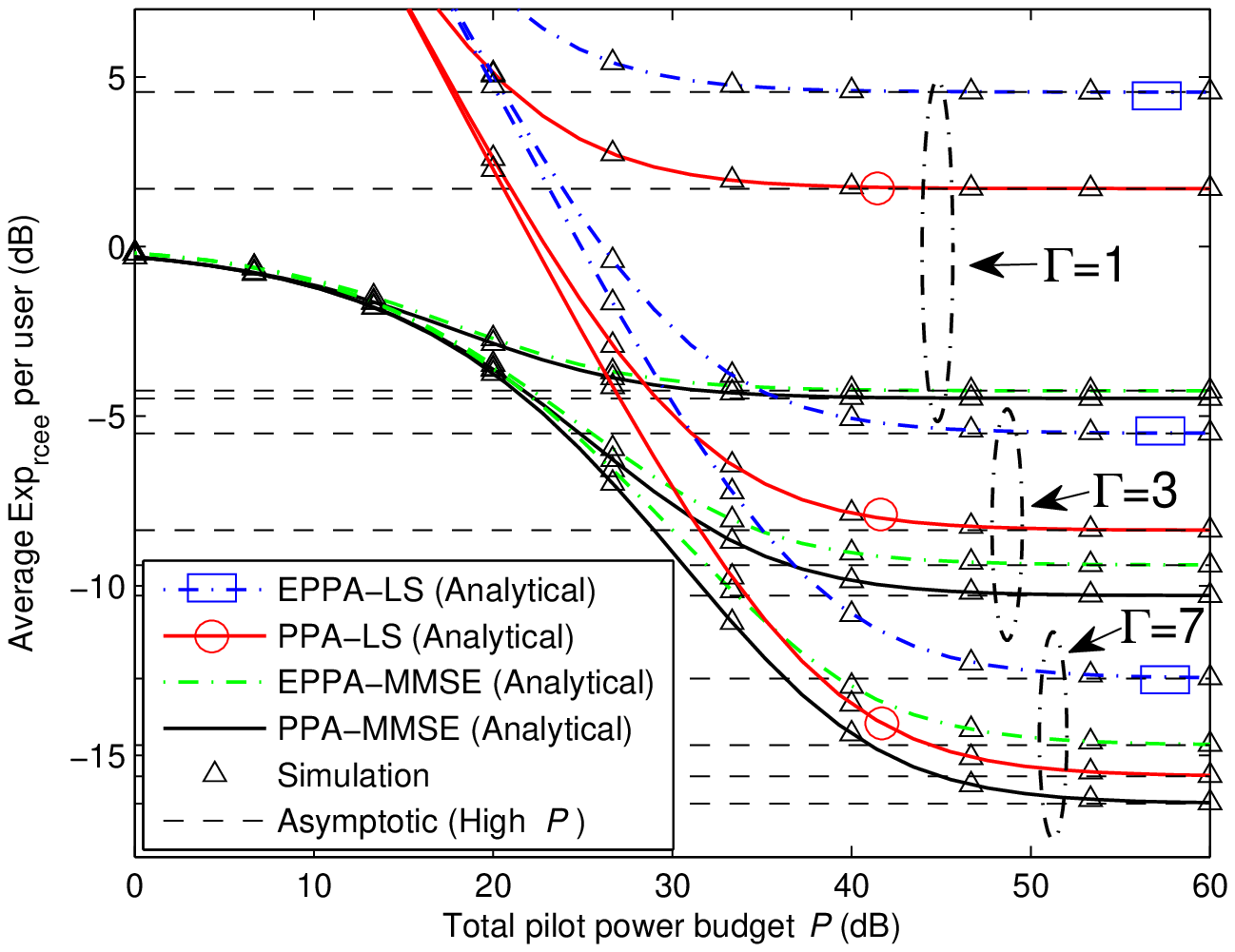}
}
\caption{The CDF of the average $\rm {{Exp}}_{ {rcee}}$ per user  for asymptotically large antenna numbers with $P=40$ dB, as well as, the average $\rm {{Exp}}_{{rcee}}$ per user as the total pilot power budget $P$ increases with  $M=200$.}
\label{fig}
\end{figure*}
\begin{figure*}[!t]
\subfigure[Minimum achievable uplink rate] { \label{fig:(a)}
\includegraphics[width=3.1in,height=2.1in]{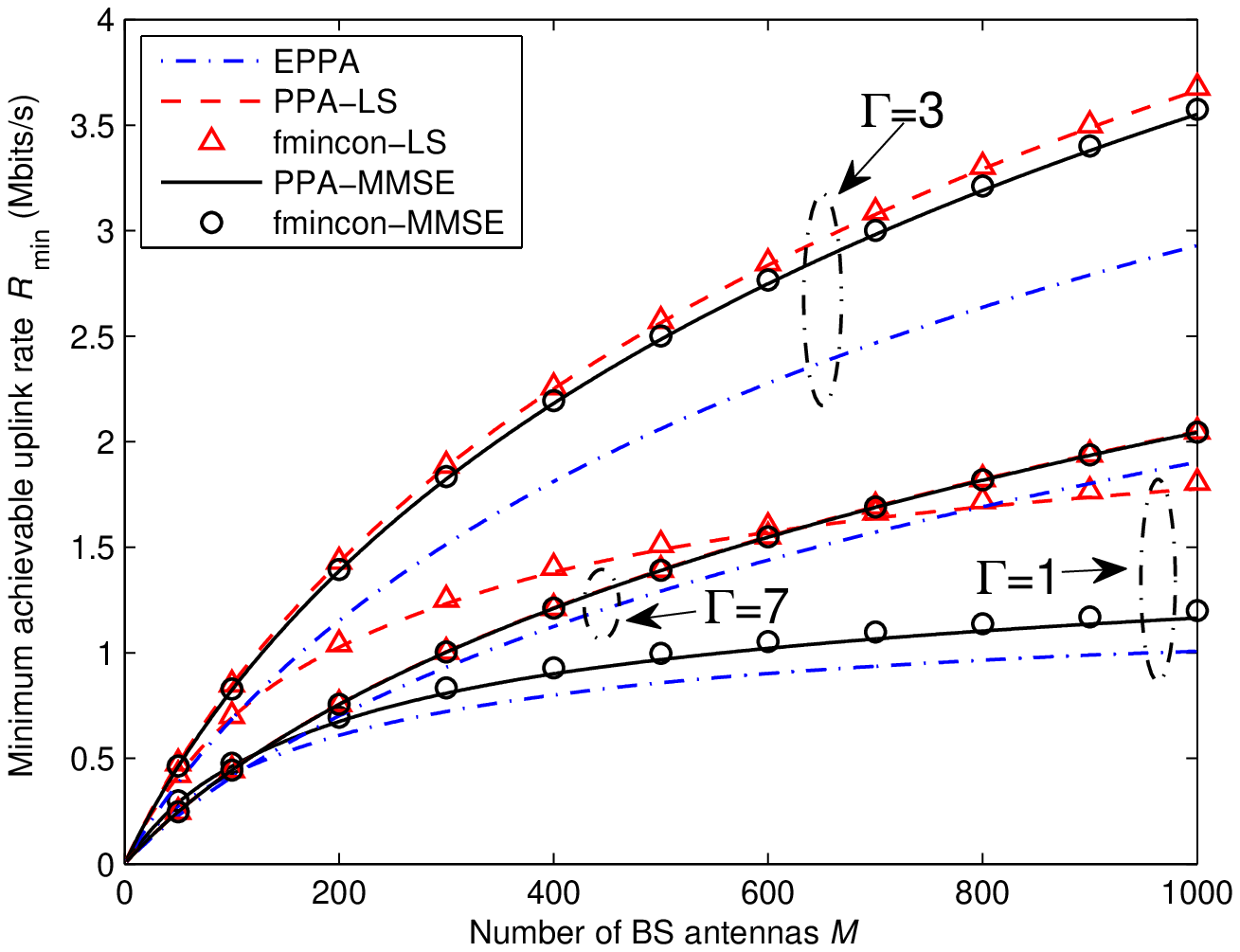}
}
\subfigure[CDF of average achievable uplink rate] { \label{fig:(b)}
\includegraphics[width=3.1in,height=2.1in]{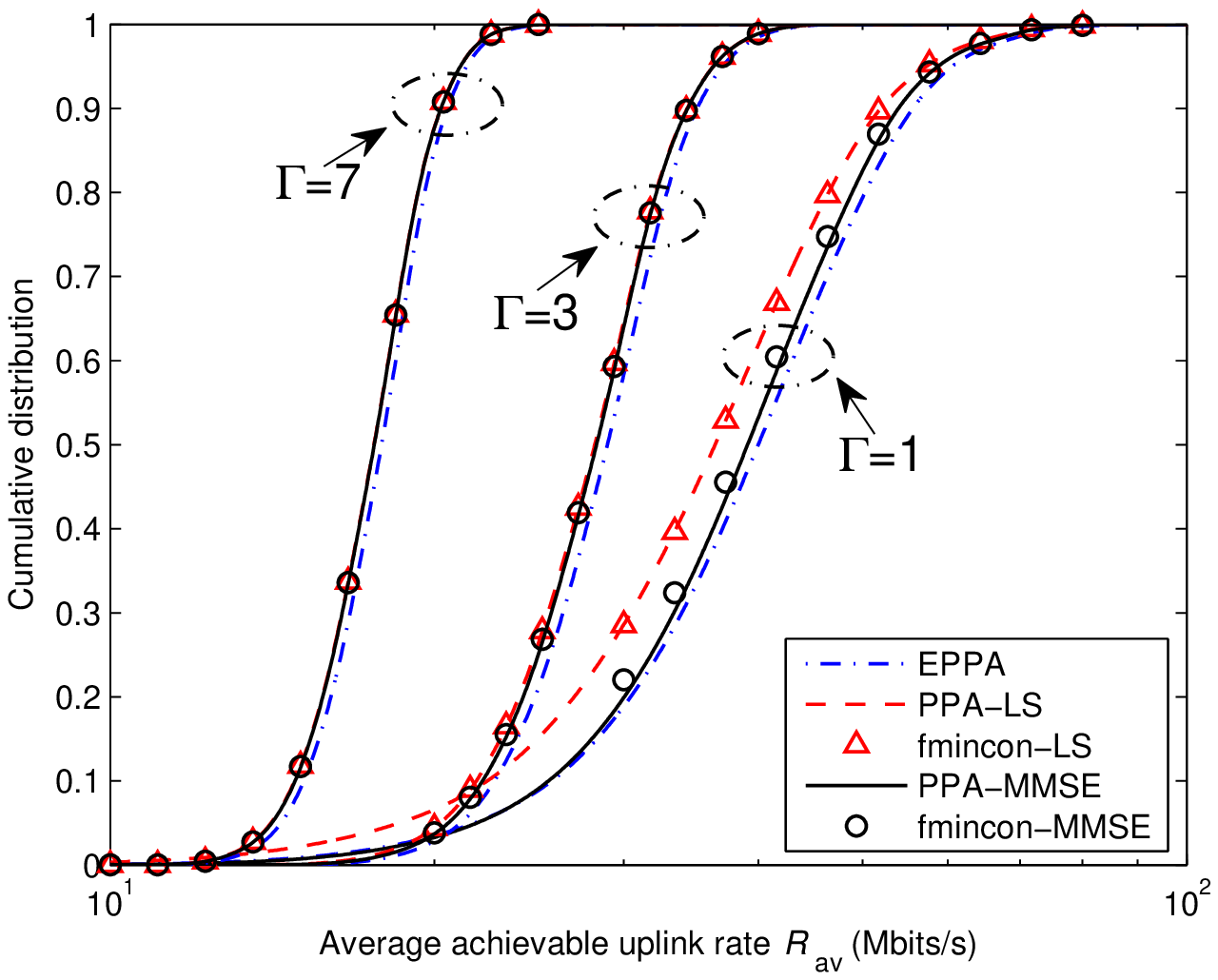}
}
\caption{The minimum achievable uplink rate as a function of the number of BS antennas with $P=40$ dB and $\rho_{u}=20$ dB, as well as, the  CDF of the average achievable uplink rate  with  asymptotically large antenna numbers and $P=40$ dB.}
\label{fig}
\end{figure*}

In this subsection, we compare the channel estimation performance of the PPA scheme and the EPPA scheme. The objective cell is again cell 1. In this cellular system, we consider $K=10$ users in each cell and set $\mu=3$. The other parameters are the same as in Section IV-A. For each analytical result, 100 independent large-scale coefficient realizations are generated. The results of the PPA scheme and the EPPA scheme are obtained by averaging over 100 independent small-scale fading channels for each large-scale channel realization. Moreover, different frequency reuse factors  are also considered as in \cite{Marzetta10}.

Fig. 3 gives the analytical and Monte-Carlo simulated average $\rm {{Exp}}_{{rcee}}$ per user with the LS and MMSE estimation methods, respectively. Results are shown for different frequency reuse factors, and $P=40$ dB. Again, we see that in all cases the ``Analytical" curves (obtained from (23)) match precisely with the simulated curves (obtained from (22)), which proves the validity of {\emph {Theorem 2}}.  When $M\rightarrow\infty$, we also see that all results tend to different constants, which match the asymptotic expressions, derived in {\emph {Corollary 3}} and {\emph {Corollary 5}}, respectively. Moreover, it is found that the PPA scheme performs systematically better than the EPPA scheme, which justifies the effectiveness of our PPA scheme. Also, considering the gap between the PPA scheme and the EPPA scheme, from Fig. 3a, as the frequency reuse factor increases, the gap between the EPPA scheme and the PPA scheme is almost fixed for LS. From Fig. 3b,  when the frequency reuse factor increases, the gap between the EPPA scheme and the PPA scheme is gradually increased for MMSE  estimation.

Fig. 4a presents the cumulative distribution function (CDF) of the analytical average $\rm { {Exp}}_{{rcee}}$ per user with $P=40$ dB and frequency reuse factors $\Gamma=1$, 3, and 7, for the LS and MMSE estimation methods, respectively, under the setting of asymptotically large antenna number. This figure shows that using the PPA algorithm to allocate pilot power  improves the average $\rm {{Exp}}_{{rcee}}$ per user in the whole probability distribution range compared with the EPPA scheme for all cases. We also see that by increasing the frequency reuse factor, the channel estimation performances of the LS and MMSE methods get closer for both the EPPA  and PPA schemes, respectively. To be more specific, when $\Gamma=1$, the co-channel interference is still  stronger than the channel power between the user in cell 1 and the BS in cell 1, so the average $\rm { {Exp}}_{ {rcee}}$ per user of the LS estimation method is much worse than the MMSE, since the former treats co-channel interference as noise and fails to deliver precise estimates. On the contrary, when $\Gamma=7$, the co-channel interference becomes small compared with the thermal noise at the receiver, such that the performance of the LS PPA scheme approaches the performance of the MMSE PPA scheme.

Fig. 4b investigates the impact of the total pilot power budget $P$ on the average $\rm { {Exp}}_{{rcee}}$ per user performance. In this figure, the number of BS antennas is set to 200. It shows that the analytical values and simulation values are almost indistinguishable for both EPPA and PPA schemes, regardless of the value of $P$. When ${P}\rightarrow\infty$, the average $\rm {{Exp}}_{{rcee}}$ per user approaches to different constant values, which match the asymptotic expressions (obtained by calculating the limit of (23) when $P\rightarrow\infty$ and combining with {\emph {Theorem 4}} and {\emph {Corollary 6}}) well, respectively. This result showcases the beneficial impact of larger frequency reuse factor on the channel estimation performance, since the average $\rm { {Exp}}_{{rcee}}$ per user is lower for larger $\Gamma$. Moreover, as the frequency reuse factor and $P$ increase, the channel estimation performance of LS and MMSE estimation methods are close to each other for both the EPPA and PPA schemes. In other words, by extending the distance between cells which use the same frequency benefits the channel estimation performance, and the simple estimation method (LS) approaches the more sophisticated estimation method (MMSE) when $P$ is big enough.

\subsection{Achievable Uplink Rate Comparison}

In this subsection, we compare the achievable uplink rate performance of the PPA scheme and the EPPA scheme. The relative parameters are the same as subsection B except for $\rho_{u}=20$ dB and $P=40$ dB.
Also, by considering the methodology of \cite{Marzetta10}, we set $T_o=71.4$ ms, $T_u=66.7$ ms, $B=20$ MHz, and $(T_s-T_p)/T_s=3/7$.
For comparison, we define two metrics called ``Minimum achievable uplink rate" and ``Average achievable uplink rate" in target cell 1, which are given as
\begin{align}
{R}_{\rm {min}}=\min\{{R}_{11},\ldots,{R}_{1k},\ldots,{R}_{1K}\},\tag{46}
\end{align}
and
\begin{align}
{R}_{\rm av}=\frac{1}{K}\sum\limits_{k=1}^{K}{R}_{1k},\tag{47}
\end{align}
respectively.

Fig. 5a gives the minimum achievable uplink rate based on (17), (19), and (46) for both the EPPA and PPA schemes, as well as, the fmincon method.\footnote{Based on {\emph {Theorem 1}}, for convenience, we only need to use the case EPPA scheme to replace the cases of the LS EPPA and MMSE EPPA schemes in Fig. 5.} For all cases, when $M$ increases, ${R}_{\rm min}$ increases and the PPA scheme performance is almost the same as the fmincon performance. To be more specific, when $\Gamma=1$, the PPA scheme performance is better than the EPPA scheme. Moreover, the LS PPA scheme is better than the MMSE PPA scheme since (19) is an increasing function of the target user's pilot power and the PPA algorithm allocates more power to the user whose relative channel estimation performance is bad; yet, note that for a given user's pilot power in the taget cell, the LS estimation performance is worse than the MMSE. As $\Gamma=3$, although the available bandwidth becomes small, the effective SINR grows larger since the co-channel interference becomes small compared with $\Gamma=1$. Hence, compared with the EPPA scheme, the PPA scheme performance improves considerably. Note that the performance of the MMSE PPA scheme approaches that of the LS  PPA scheme. When $\Gamma=7$, compared with $\Gamma=3$, the available bandwidth becomes substantially smaller, while the effective SINR grows marginally. Hence, regarding ${R}_{\rm min}$, $\Gamma=7$ performs worse than $\Gamma=3$.  Also, now the MMSE performance is almost the same that of the LS. In summary, compared with the EPPA scheme, for both the LS and MMSE estimation methods, our PPA scheme improves substantially the performance of the user with the minimum achievable uplink rate.

Fig. 5b gives the CDF of the average achievable uplink rate based on  (17), (19), and (47) for both the EPPA and PPA schemes, as well as, the fmincon method. For all cases, the performance difference between the PPA scheme and the fmincon is almost negligible. For the case of the LS estimation method, when $\Gamma=1$, the LS PPA scheme performance is slightly worse that of the EPPA scheme since the proposed PPA algorithm puts more pilot power to the user who suffers strong co-channel interference; this affects negatively the rate performance of the users who suffer small co-channel interference. As $\Gamma=3$ and $\Gamma=7$, the difference of co-channel interference level between the users in the cell becomes small; hence, the difference between the PPA scheme and the EPPA scheme can be ignored.
For the case of the MMSE estimation method, for all frequency reuse factors, the PPA scheme is almost the same as the EPPA scheme. Moreover, with an increasing frequency reuse factor, the performance of the LS approaches the performance of the MMSE. In summary, compared with the EPPA scheme, for both the LS and MMSE estimation methods, our PPA scheme offers almost the same average achievable uplink rate.

\section{Conclusion}
In this paper, we studied the performance of channel estimation and achievable uplink rate for multi-cell massive MIMO systems. We provided a modified NMSE metric called RCEE and deduced new, tractable, closed-form expressions for $\rm {Exp}_{rcee}$ and achievable uplink rate for the LS and MMSE estimation methods, respectively. We found that RCEE and $\rm {Exp}_{rcee}$ tend to the same constant when $M\rightarrow\infty$, which reflects the channel hardening effect. Due to the obtained closed-form expressions for $\rm {Exp}_{rcee}$, we proposed a PPA algorithm to minimize the average $\rm { {Exp}}_{{rcee}}$ per user with a total pilot power budget. The channel estimation performance of the PPA scheme and EPPA scheme were compared for both the LS and MMSE estimation methods. Regarding the differences between the PPA scheme and EPPA scheme, the LS gap remains almost fixed irrespective of the frequency reuse factor, whilst the MMSE gap increases. Moreover, when co-channel interference is small and the pilot power is high enough, the LS scheme channel estimation performance approaches the MMSE scheme performance. Hence, from a design point of view, the simple LS estimation method is a very viable choice. Finally, the average achievable uplink rate of the PPA scheme is almost the same as the EPPA scheme, while, compared with the EPPA scheme, the PPA scheme improves significantly the minimum achievable uplink rate.

\appendices
\section{Proof of Theorem 1}
To evaluate the ${\text {SINR}}_{jk}$ in (18), we define four terms
\begin{align}
\mathfrak{A}&\triangleq\left|\mathbb{E}\left\{\hat{\bf h}_{jjk}^{\text H}{\bf h}_{jjk}\right\}\right|^{2},\tag{48}\\
\mathfrak{B}&\triangleq\mathbb{E}\left\{\left|\hat{\bf h}_{jjk}^{\text H}{\bf h}_{jln}\right|^{2}\right\}, (n\neq k),\tag{49}\\
\mathfrak{C}&\triangleq\mathbb{E}\left\{\left|\hat{\bf h}_{jjk}^{\text H}{\bf h}_{jlk}\right|^{2}\right\}, \tag{50}\\
\mathfrak{D}&\triangleq\mathbb{E}\left\{\left\|\hat{\bf h}_{jjk}\right\|_{2}^{2}\right\}.\tag{51}
\end{align}

Although $\hat{\bf h}_{jjk}$ has different expressions for the LS and
MMSE estimation methods, the corresponding proofs for ${\text {SINR}}_{jk}$ are similar. Hence,
it is convenient to only study the case of the LS  estimation.
\begin{itemize}
    \item Calculate $\mathfrak{A}$: Substituting (3) and (9) into (48), we get
    \begin{align}
     \mathfrak{A}=\left|\mathbb{E}\left\{\left\|{\bf h}_{jjk}\right\|_{2}^{2}\right\}\right|^{2}=M^{2}\beta_{jjk}^{2}.\tag{52}
    \end{align}

    \item Calculate $\mathfrak{B}$:  By considering the $\hat{\bf h}_{jjk}^{\text {LS}}$ is uncorrelated with ${\bf h}_{jln}$ when $k\neq n$,  we can obtain
     \begin{align}
      \mathfrak{B}=\mathbb{E}\left\{(\hat{\bf h}_{jjk}^{\text {LS}})^{\text H}\mathbb{E}\left\{{\bf h}_{jln}{\bf h}_{jln}^{\text H}\right\}\hat{\bf h}_{jjk}^{\text {LS}}\right\}.\tag{53}
    \end{align}
    Then, substituting (3) and (9) into (53), yields
    \begin{align}
      \mathfrak{B}=\frac{M\beta_{jln}}{\rho_{jk}}\left(\sum\limits_{l=1}^{L}\rho_{lk}\beta_{jlk}+1\right).\tag{54}
    \end{align}

    \item Calculate $\mathfrak{C}$: Since $\hat{\bf h}_{jjk}^{\text {LS}}$ is correlated with ${\bf h}_{jlk}$, substituting (9) into (50), we can get
    \begin{align}
      \mathfrak{C}&=\frac{1}{\rho_{jk}}\bigg(\underbrace{\mathbb{E}\left\{\left|\left(\sum\limits_{c\neq l}^{L}\sqrt{\rho_{ck}}{\bf h}_{jck}+{\bf N}_{j}{\underline{{\bf s}}}_{jk}\right)^{\text H}{\bf h}_{jlk}\right|^{2}\right\}}_{{C_1}}+\rho_{lk}\underbrace{\mathbb{E}\left\{\|{\bf h}_{jlk}\|_{2}^{4}\right\}}_{ {C_2}}\bigg),\tag{55}
    \end{align}
    where the closed-form expression for $C_1$ can be obtained based on the similar way for obtaining (49), that is,
    \begin{align}
      {{C_1}}=M\beta_{jlk}\left(\sum\limits_{c\neq l}^{L}\rho_{ck}\beta_{jck}+1\right),\tag{56}
    \end{align}
    whilst for the term $C_2$, using the properties of Wishart matrices \cite[{\emph {Lemma 2.9}}]{Tulino04}, $C_2$ is given by
    \begin{align}
          {{C_2}}=M(M+1)\beta_{jlk}^{2}.\tag{57}
    \end{align}
    Therefore, substituting (56) and (57) into (55) and simplifying, we can write $ \mathfrak{C}$ as follows
    \begin{align}
      \mathfrak{C}=\frac{1}{\rho_{jk}}\!\!\left(\!\!M\beta_{jlk}\!\!\left(\sum\limits_{c=1}^{L}\!\rho_{ck}\beta_{jck}\!\!+\!\!1\!\!\right)\!\!+\!\!M^{2}\rho_{lk}\beta_{jlk}\!\!\right).\!\!\!\tag{58}
    \end{align}

    \item Calculate $\mathfrak{D}$: With the help of (3) and (9), $\mathfrak{D}$ is given by
    \begin{align}
      \mathfrak{D}=\frac{M}{\rho_{jk}}\left(\sum\limits_{l=1}^{L}\rho_{lk}\beta_{jlk}+1\right).\tag{59}
    \end{align}
\end{itemize}
Finally, substituting (52), (54), (58), and (59) into (18) and simplifying, the closed-form expression for ${\text {SINR}}_{jk}$ based on LS estimation method is obtained.
\hfill\rule{3mm}{3mm}

\section{Proof of Theorem 2}
Although ${\Delta}_{jk}$ has different closed-form expressions for LS and
MMSE estimation methods, the corresponding proofs are similar. Hence,
it is convenient to only study the case of the LS estimation. By the definition of conditional expectation, from (21) and (22), we obtain
\begin{align}
{\Delta}_{jk}=\mathbb{E}\left\{\mathbb{E}\left\{\underbrace{\left\|{\bf h}_{jjk}-\hat{{\bf h}}_{jjk}^{\text {LS}}\right\|_{2}^{2}\bigg|{\bf h}_{jjk}}_{I_1}\right\}\underbrace{\frac{1}{\left\|{\bf h}_{jjk}\right\|_{2}^{2}}}_{I_2}\right\},\tag{60}
\end{align}
where $\mathbb{E}\{I_1\}$ is called the conditional expectation of $\|{\bf h}_{jjk}-\hat{{\bf h}}_{jjk}^{\text {LS}}\|_{2}^{2}$ given ${\bf h}_{jjk}$. To this end, we substitute (9) into $I_{1}$. Then, we get
\begin{align}
\mathbb{E}\{I_1\}=\mathbb{E}\left\{\left\|\sum_{l\neq j}^{L}\frac{\sqrt {{\rho}_{lk}}}{\sqrt {{\rho}_{jk}}}{\bf h}_{jlk}+\frac{1}{\sqrt {{\rho}_{jk}}}{\bf N}_{j}{\underline{{\bf s}}}_{jk}\right\|_{2}^{2}{\bigg|}{\bf h}_{jjk}\right\}.\tag{61}
\end{align}
It is obvious that (61) does not depend on ${\bf h}_{jjk}$. Moreover, based on the properties of the expectation operator and the trace operator and the assumption of the user's channel, we evaluate (61) to yield
\begin{align}
\mathbb{E}\{I_1\}\!=\!{\rm {tr}}\!\left(\!\mathbb{E}\left\{\sum\limits_{l\neq j}^{L}\frac{\rho_{lk}}{\rho_{jk}}{\bf h}_{jlk}{\bf h}_{jlk}^{\text H}+\frac{1}{\rho_{jk}}{\bf N}_{j}{\underline{{\bf s}}}_{jk}\left({\bf N}_{j}{\underline{{\bf s}}}_{jk}\right)^{\text H}\right\}\!\right)\!.\tag{62}
\end{align}
Then, substituting (3) into (62), we have the following result
\begin{align}
\mathbb{E}\{I_1\}=\frac{M}{\rho_{jk}}\left(\sum\limits_{l\neq j}^{L}\rho_{lk}\beta_{jlk}+1\right).\tag{63}
\end{align}

To evaluate $I_{2}$ in (60), we firstly consider the case of $M=1$. Hence, we denote $I_{2}$ as $I_2^{M=1}$. Moreover, ${h}_{jjk}$ is now a scalar stochastic quantity $\sim\mathcal {C}\mathcal {N}(0, \beta_{jjk})$. Based on the special structure of $I_2^{M=1}$, after some manipulations, it is easy to obtain
\begin{align}
\mathbb{E}\{I_2^{M=1}\}=\infty.\tag{64}
\end{align}
When $M\geq 2$, we denote $I_2$ as $I_2^{M\geq2}$. Using the properties of Wishart matrices \cite[{\emph {Lemma 2.10}}]{Tulino04}, thus, $\mathbb{E}\{I_2^{M\geq2}\}$ is reduced to
\begin{align}
    \mathbb{E}\{I_2^{M\geq2}\}=\frac{1}{\beta_{jjk}(M-1)}.\tag{65}
\end{align}
Hence, by substituting (63)-(65) into (60) and simplifying, the closed-form expression for ${\Delta}_{jk}$ based on LS estimation method is obtained. \hfill\rule{3mm}{3mm}

\section{Proof of Theorem 4}
In this proof, the main challenge is to obtain the user groups
${\mathcal {K}}_{{P},\rm min}$, ${\mathcal {K}}_{{P},\rm max}$ and ${\mathcal {K}}_{P}$ in {\bf {Algorithm 1}} when $P\rightarrow\infty$. However, $P\rightarrow\infty$ means that the thermal noise in the BS can be ignored. Hence, inspired by (29), we formulate a new optimization problem by replacing the ${\Delta}_{jk}$ in (27) with ${\dot{\Delta}}_{jk}$ and its subproblem by replacing the ${\tilde{\Delta}}_{jk}$ in (29) with ${\dot{\Delta}}_{jk}$, where
\begin{align}
\begin{split}
    {\dot{\Delta}}_{jk}\!\triangleq\!\left\{\!
    \begin{array}{*{20}l}
   \frac{M\sum\limits_{l\neq j}^{L}{\dot{\rho}_{lk}}{\beta_{jlk}}}{\left(M-1\right){\dot{\rho}_{jk}}{\beta_{jjk}}}, &{\text {LS}},\\
    \\
    \frac{M\sum\limits_{l\neq j}^{L}{\dot{\rho}_{lk}}{\beta_{jlk}}}{\left(M-1\right)\sum\limits_{l=1}^{L}{\dot{\rho}_{lk}}{\beta_{jlk}}}, &{\text {MMSE}}.
     \end{array}\right.
 \end{split}\tag{66}
\end{align}
Using the Lagrange multiplier method \cite{Boyd04}, we can determine the optimal solution $\dot{\rho}_{jk}^{\ast}$ of this new optimization problem, which is similar with the format of (31)-(34) that we omit it.
Moreover, we find that ${\dot{\rho}_{jk}^{\ast}}$ is proportional to $P$ for the LS and MMSE estimation methods, respectively.
Therefore, since the PPA algorithm  solves (27) with the help of (29), we also use the core idea of PPA algorithm to solve this new optimization problem with the help of its subproblem. It shows that the output user groups ${\dot{\mathcal {K}}}_{{P},\rm min}$, $\dot{\mathcal {K}}_{{P},\rm max}$, and $\dot{\mathcal {K}}_{P}$,\footnote{To avoid confusion, we redefine these three user groups when using the core idea of PPA algorithm to solve this new optimization problem.} whose cardinalities are ${\dot{ {K}}}_{{P},\rm min}$, $\dot{{K}}_{{P},\rm max}$, and $\dot{ {K}}_{P}$, respectively,  will not depend on $P$. That is, $\forall P_1, P_2>0$, we have
$\dot{\mathcal {K}}_{{P_1},\rm min}=\dot{\mathcal {K}}_{{P_2},\rm min},\  \dot{\mathcal {K}}_{{P_1},\rm max}=\dot{\mathcal {K}}_{{P_2},\rm max}, \ \text{and}\  \dot{\mathcal {K}}_{{P_1}}=\dot{\mathcal {K}}_{{P_2}}$.
Hence, we use ${\dot{\mathcal {K}}}_{\rm min}$, $\dot{\mathcal {K}}_{\rm max}$, and $\dot{\mathcal {K}}_{}$ to replace ${\dot{\mathcal {K}}}_{{P},\rm min}$, $\dot{\mathcal {K}}_{{P},\rm max}$, and $\dot{\mathcal {K}}_{{P}}$, respectively. Also, the ${\dot{ {K}}}_{\rm min}$, $\dot{{K}}_{\rm max}$, and $\dot{ {K}}_{}$ are obtained for the same reason.
Then, we use ${\dot{\mathcal {K}}}_{\rm min}$, $\dot{\mathcal {K}}_{\rm max}$, and $\dot{\mathcal {K}}_{}$ to approximate the user groups ${{\mathcal {K}}}_{{P},\rm min}$, ${\mathcal {K}}_{{P},\rm max}$, and ${\mathcal {K}}_{P}$, respectively, when $P\rightarrow\infty$.
Later, considering ${\emph {Corollary 3}}$, we hold $M\rightarrow\infty$ and study the channel estimation performance when $P\rightarrow\infty$.
Moreover, it is obvious that the first and second expressions in both (37) and (38) are easy to obtain with the help of the limit theory. Therefore, this proof focuses on the third expressions in  both (37) and (38).

Although, when $M\rightarrow\infty$ and $P\rightarrow\infty$, the limit value of ${\Delta}_{jk}$ has different closed-form expressions for LS and MMSE estimation
methods, the corresponding proofs are similar. Hence,
it is convenient to only study the case of the LS  estimation.
We now substitute the first line of (31) into the first line of (25) and considering that the user groups ${\mathcal {K}}_{{P}}$, ${\mathcal {K}}_{{P},{\rm min}}$, and ${\mathcal {K}}_{{P},{\rm max}}$ are obtained by the PPA algorithm; hence, for the users in groups ${\mathcal {K}}_{{P}}$, we replace $\sum\limits_{k=1}^{K}$ with $\sum\limits_{k\in {\mathcal {K}}_{P}}$ and $P$ with $P-{\rho_{\rm min}}{{K}}_{P, \rm min}-{\rho_{\rm max}}{{K}}_{P, \rm max}$ and set $\alpha={\rho_{\rm min}}K/{P}$, as well as, we consider the definition of $\mu$, to obtain
\begin{align}
\frac{\sum\limits_{l\neq j}^{L}{\rho_{lk}}{\beta_{jlk}}+1}{\rho_{jk}\beta_{jjk}}&=\frac{\bigg({\sum\limits_{l\neq j}^{L}{\delta_{lk}{P}}{\beta_{jlk}}+1}\bigg)\sum\limits_{k\in {\mathcal {K}}_{P}}\bigg(\frac{w_{jk}}{\beta_{jjk}} \bigg({\sum\limits_{l\neq j}^{L}{\delta_{lk}{P}}{\beta_{jlk}}+1}\bigg)\bigg)^{\frac{1}{2}}}{\bigg(P-{{K}}_{P, \rm min}\frac{\alpha{P}}{K}-{{K}}_{P, \rm max}\frac{\mu{P}}{K}\bigg)\bigg({w_{jk}}{\beta_{jjk}} \bigg({\sum\limits_{l\neq j}^{L}{\delta_{lk}{P}}{\beta_{jlk}}+1}\bigg)\bigg)^{\frac{1}{2}}}.\tag{67}
\end{align}
Finally, using the limit theory when $P\rightarrow\infty$, and simplifying, we obtain the third expression in (37).\hfill\rule{3mm}{3mm}

\ifCLASSOPTIONcaptionsoff
  \newpage
\fi

\footnotesize
\bibliographystyle{ieee}
\bibliography{CASSreference}

\begin{thebibliography}{99}
\bibitem{Marzetta10}
T. L. Marzetta, ``Noncooperative cellular wireless with
unlimited numbers of base station antennas," \textit { IEEE Trans. Wireless Commun.}, vol. 9, no. 11, pp. 3590-3600, Nov. 2010.


\bibitem{Rusek13}
F. Rusek, D. Persson, B. K. Lau, E. G. Larsson, T. L. Marzetta, O. Edfors, and F. Tufvesson, ``Scaling up MIMO: Opportunities and challenges with very large arrays," \textit {IEEE Signal Process. Mag.}, vol. 30, no. 1, pp. 40-60, Jan. 2013.

\bibitem{QiZhang14}
Q. Zhang, S. Jin, K.-K. Wong, H. Zhu, and M. Matthaiou, ``Power scaling of uplink massive MIMO systems with arbitrary-rank channel means," {\textit {IEEE J. Sel. Topics Signal Process.}}, vol. 8, no. 5, pp. 966-981, Oct. 2014.




\bibitem{Boccardi14}
F. Boccardi, R. W. Heath Jr., A. Lozano, T. L. Marzetta, and P. Popovski,
``Five disruptive technology directions for 5G," \textit {IEEE Commun. Mag.},
vol. 52, no. 2, pp. 74-80, Feb. 2014.


\bibitem{Ngo13}
H. Q. Ngo, E. G. Larsson, and T. L. Marzetta, ``Energy and spectral efficiency of
very large multiuser MIMO systems," \textit {IEEE Trans. Commun.}, vol. 61, no. 4, pp. 1436-1449, Apr. 2013.

\bibitem{Jose11}
J. Jose, A. Ashikhmin, T. L. Marzetta, and S. Vishwanath, ``Pilot contamination and precoding in
multi-cell TDD systems," \textit {IEEE Trans. Wireless Commun.}, vol. 10, no. 8, pp. 2640-2651, Aug. 2011.

\bibitem{Hoydis13}
J. Hoydis, S. ten Brink, and M. Debbah, ``Massive MIMO in the UL/DL of cellular networks: How many antennas do we need?," \textit {IEEE J. Sel.
Areas Commun.}, vol. 31, no. 2, pp. 160-171, Feb. 2013.



\bibitem{Bjornson14}
E. Bj{\"{o}}rnson, J. Hoydis, M. Kountouris, and M. Debbah, ``Massive MIMO systems with non-ideal hardware:
Energy efficiency, estimation, and capacity limits," \textit { IEEE Trans. Inf. Theory}, vol. 60, no. 11, pp. 7112-7139, Nov. 2014.


\bibitem{Yin13}
H. Yin, D. Gesbert, M. Filippou, and Y. Liu, ``A coordinated approach to channel estimation in
large-scale multiple-antenna systems," \textit {IEEE J. Sel. Areas Commun.}, vol. 31, no. 2, pp. 264-273, Feb. 2013.


\bibitem{JiankangZhang14}
J. Zhang, B. Zhang, S. Chen, X. Mu, M. El-Hajjar, and L. Hanzo, ``Pilot contamination elimination for large-scale
multiple-antenna aided OFDM systems," \textit {IEEE J. Sel. Topics
Signal Process.}, vol. 8, no. 5, pp. 759-772, Oct. 2014.

\bibitem{Marzetta15}
T. L. Marzetta, ``Massive MIMO: An introduction," \textit {Bell
Labs Tech. J.}, vol. 20, pp. 11-22, 2015.



\bibitem{Ngo12}
H. Q. Ngo and E. G. Larsson, ``EVD-based channel estimation in multicell multiuser MIMO systems with very large antenna arrays," in {\textit {Proc. IEEE ICASSP,}}  Mar. 2012, pp. 3249-3252.


\bibitem{Muller14}
R. R. M{\"{u}}ller, L. Cottatellucci, and M. Vehkaper{\"{a}}, ``Blind pilot decontamination," \textit {IEEE J. Sel. Topics Signal Process.}, vol. 8, no. 5, pp. 773-786, Oct. 2014.

\bibitem{Ma14}
J. Ma and L. Ping, ``Data-aided channel estimation in
large antenna systems," \textit {IEEE Trans. Signal Process.}, vol. 62, no. 12, pp. 3111-3124, Jun. 2014.


\bibitem{Wen15}
C.-K. Wen, S. Jin, K.-K. Wong, J.-C. Chen, and P. Ting, ``Channel estimation for massive MIMO using
Gaussian-mixture Bayesian learning," \textit { IEEE Trans. Wireless Commun.}, vol. 14, no. 3, pp. 1356-1368, Mar. 2015.

\bibitem{Shariati14}
N. Shariati, E. Bj{\"{o}}rnson, M. Bengtsson, and M. Debbah, ``Low-complexity polynomial channel estimation in large-scale MIMO with arbitrary statistics," \textit {IEEE J. Sel. Topics Signal Process.}, vol. 8, no. 5, pp. 815-830, Oct. 2014.

\bibitem{Hu15}
D. Hu, L. He, and X. Wang, ``Semi-blind pilot decontamination for massive
MIMO systems," \textit {IEEE Trans. Wireless Commun.}, vol. 15, no. 1, pp. 525-536, Jan. 2016.

\bibitem{Zhu15}
X. Zhu, Z. Wang, C. Qian, L. Dai, J. Chen, S. Chen, and L. Hanzo, ``Soft pilot reuse and multi-cell block diagonalization precoding for
massive MIMO systems," \textit {IEEE Trans. Veh. Technol.}, 2015, to appear. [Online]. Available: http://arxiv.org/abs/1507.04213

\bibitem{Ngo14}
H. Q. Ngo, M. Matthaiou, and E. G. Larsson, ``Massive MIMO with optimal power and training duration allocation," \textit {IEEE Wireless Commun. Lett.}, vol. 3. no. 6, pp. 605-608, Dec. 2014.

\bibitem{Guo14}
K. Guo, Y. Guo, G. Fodor, and G. Ascheid, ``Uplink power control with MMSE receiver in multi-cell MU-massive-MIMO systems," in \textit {Proc. IEEE ICC},  Jun. 2014, pp. 5184-5190.


\bibitem{Xiang14}
Z. Xiang, M. Tao, and X. Wang, ``Massive MIMO multicasting in noncooperative
cellular networks," \textit {IEEE J. Sel. Areas Commun.}, vol. 32, no. 6, pp. 1180-1193, Jun. 2014.

\bibitem{Zheng14}
X. Zheng, H. Zhang, W. Xu, and X. You, ``Optimized pilot power allocation for heterogeneous users in massive MIMO downlinks," in \textit {Proc. IEEE ICTC},  Oct. 2014, pp. 246-250.

\bibitem{Saxena15}
V. Saxena, G. Fodor,  and E. Karipidis, ``Mitigating pilot contamination by pilot reuse and power control schemes for massive MIMO systems," in \textit {Proc. IEEE VTC Spring},  May 2015.

\bibitem{Cheng15}
H. V. Cheng,  E. Bj{\"{o}}rnson,  and E. G. Larsson, ``Uplink pilot and data power control for single cell massive MIMO systems with MRC," in \textit {Proc. IEEE ISWCS},  Aug. 2015. [Online]. Available: http://arxiv.org/abs/1509.02633


\bibitem{Kay93}
S. M. Kay, \textit {Fundamentals of Statistical Signal Processing: Estimation
Theory}. Prentice-Hall, Inc. Upper Saddle River, NJ, USA, 1993.


\bibitem{Bjornson15}
E. Bj\"{o}rnson, E. G. Larsson, and  M. Debbah, ``Massive MIMO for maximal spectral efficiency: How many users and pilots should be allocated?," \textit { IEEE Trans. Wireless Commun.}, vol. 15, no. 2, pp. 1293-1308, Feb. 2016.

\bibitem{Neumann15}
D. Neumann, M. Joham, and W. Utschick, ``Channel estimation in massive MIMO systems," Mar. 2015. [Online]. Available: http://arxiv.org/abs/1503.08691

\bibitem{Yin14}
H. Yin, D. Gesbert, and L. Cottatellucci, ``Dealing with interference in distributed large-scale
MIMO systems: A statistical approach," {\textit {IEEE J. Sel. Topics Signal Process.}}, vol. 8, no. 5, pp. 942-953, Oct. 2014.



\bibitem{Ngo132}
H. Q. Ngo, E. G. Larsson, and T. L. Marzetta, ``The multicell multiuser MIMO uplink with very large antenna arrays and a finite-dimensional channel," \textit {IEEE Trans. Commun.}, vol. 61, no. 6, pp. 2350-2361, Jun. 2013.




\bibitem{Narasimhan14}
T. L. Narasimhan and A. Chockalingam, ``Channel hardening-exploiting message passing
(CHEMP) receiver in large-scale MIMO systems," \textit {IEEE J. Sel. Topics
Signal Process.}, vol. 8, no. 5, pp. 847-860, Oct. 2014.


\bibitem{Boyd04}
S. Boyd and L. Vandenberghe, \textit {Convex Optimization}. Cambridge, U.K.: Cambridge
Univ. Press, 2004.


\bibitem{Ngo141}
H. Q. Ngo, H. A. Suraweera, M. Matthaiou, and E. G. Larsson, ``Multipair full-duplex relaying with massive arrays
and linear processing," \textit {IEEE J. Sel. Areas Commun.}, vol. 32, no. 9, pp. 1721-1737, Sept. 2014.



\bibitem{Tulino04}
A. M. Tulino and S. Verd\'{u}, ``Random matrix theory and wireless
communications," \textit {Foundations Trends Commun. Inf. Theory}, vol. 1,
no. 1, pp. 1-182, Jun. 2004.

\end{thebibliography}

\end{document}